\documentclass[acmsmall,natbib,screen=true,]{acmart} 

\usepackage{booktabs} 
\usepackage[skip=0pt]{caption}
\usepackage{multicol}
\usepackage{multirow}
\usepackage[inline]{enumitem}
\usepackage{bbm}
\usepackage{acronym}
\usepackage{algorithm}
\usepackage{algorithmic}
\usepackage{array}
\usepackage{subfigure}

\AtBeginDocument{%
  \providecommand\BibTeX{{%
    \normalfont B\kern-0.5em{\scshape i\kern-0.25em b}\kern-0.8em\TeX}}}
\acrodef{SR}{sequential recommendation}
\acrodef{DL}{deep learning}
\acrodef{CV}{computer vision}
\acrodef{NLP}{neural language processing}
\acrodef{RNN}{recurrent neural network}
\acrodef{CNN}{convolutional neural network}
\acrodef{GNN}{graph neural network}
\acrodef{RL}{reinforcement learning}
\acrodef{SSL}{self-supervised learning}
\acrodef{MIP}{masked item prediction}
\acrodef{ICL}{item contrastive learning}
\acrodef{MSE}{mean squared error}
\acrodef{SSE}{stochastic shared embeddings}

\newcommand{\OurMethod}{MrTransformer}

\newcommand{\my}[1]{\textcolor{blue}{\textbf{[#1]}}}

\newcommand{\squeeze}{\vspace*{-2mm}}

\allowdisplaybreaks

\setcopyright{acmcopyright}
\copyrightyear{2021}
\acmYear{2021}
\acmDOI{xx.xxxx/xxxxxxx.xxxxxxx}
\acmPrice{15.00}
\acmISBN{978-1-4503-9999-9/18/06}
\acmJournal{TOIS}

\author{Muyang Ma}
\orcid{}
\affiliation{%
\institution{Shandong University}
\city{Qingdao}
\country{China}
}
\email{muyang0331@gmail.com}

\author{Pengjie Ren$^{*}$}
\orcid{}
\affiliation{%
\institution{Shandong University}
\city{Qingdao}
\country{China}
}
\email{jay.ren@outlook.com}

\author{Zhumin Chen}
\orcid{}
\affiliation{%
\institution{Shandong University}
\city{Qingdao}
\country{China}
}
\email{chenzhumin@sdu.edu.cn}

\author{Zhaochun Ren}
\orcid{}
\affiliation{%
\institution{Shandong University}
\city{Qingdao}
\country{China}
}
\email{zhaochun.ren@sdu.edu.cn}

\author{Huasheng Liang}
\orcid{}
\affiliation{%
\institution{WeChat, Tencent}
\city{Shenzhen}
\country{China}
}
\email{watsonliang@tencent.com}

\author{Jun Ma}
\orcid{}
\affiliation{%
\institution{Shandong University}
\city{Qingdao}
\country{China}
}
\email{majun@sdu.edu.cn}

\author{Maarten de Rijke}
\orcid{0000-0002-1086-0202}
\affiliation{
 \institution{University of Amsterdam and Ahold Delhaize}
 \city{Amsterdam and Zaandam}
 \country{The Netherlands}
}

\email{m.derijke@uva.nl}

\thanks{$^*$Corresponding author.}

\begin{document}
 
\title[Improving Transformer-based Sequential Recommenders through Preference Editing]{Improving Transformer-based Sequential Recommenders\\ through Preference Editing}


\begin{abstract}
One of the key challenges in \ac{SR} is how to extract and represent user preferences.
Traditional \ac{SR} methods rely on the next item as the supervision signal to guide preference extraction and representation.
We propose a novel learning strategy, named \textit{preference editing}.
The idea is to force the \ac{SR} model to discriminate the common and unique preferences in different sequences of interactions between users and the recommender system.
By doing so, the \ac{SR} model is able to learn how to identify common and unique user preferences, and thereby do better user preference extraction and representation.
We propose a transformer based \ac{SR} model, named \OurMethod{} (\textbf{M}ulti-p\textbf{r}eference \textbf{Transformer}), that concatenates some special tokens in front of the sequence to represent multiple user preferences and makes sure they capture different aspects through a preference coverage mechanism.
Then, we devise a \textit{preference editing}-based \acl{SSL} mechanism for training \OurMethod{} that contains two main operations: \textit{preference separation} and \textit{preference recombination}.
The former separates the common and unique user preferences for a given pair of sequences.
The latter swaps the common preferences to obtain recombined user preferences for each sequence.
Based on the preference separation and preference recombination operations, we define two types of \acl{SSL} loss that require that the recombined preferences are similar to the original ones, and that the common preferences are close to each other.

We carry out extensive experiments on two benchmark datasets.
\OurMethod{} with preference editing significantly outperforms state-of-the-art \ac{SR} methods in terms of Recall, MRR and NDCG.
We find that long sequences whose user preferences are harder to extract and represent benefit most from preference editing.
\end{abstract}



\maketitle

\acresetall


\section{Introduction}

Sequential recommendation (SR\acused{SR}) methods aim to predict the next item that the user is most likely to interact with based on his/her past interactions, such as clicking on products or watching movies.
One of the key challenges faced by \ac{SR} approaches is \textit{extract and represent user preferences} historical interaction sequences~\cite{yuan2020future,sun2020dual}.
Traditional methods for \ac{SR} rely solely on predicting the next item to guide user preference extraction and representation~\cite{donkers2017sequential,cross-domain_novelty-seeking_sr,SRGNN,pan2020rethinking,pan2020intent,ren2020nlp4rec}.
So far, correlations between interactions have not been well investigated as a supervision signal that can guide user preference extraction and representation.

Self-supervised learning (SSL\acused{SSL}) can automatically generate a supervision signal to learn representations of the data or to automatically label a dataset.
\ac{SSL} has made great progress in \ac{CV}~\cite{cv_zhai2019,cv_sun2019,cv_vandenOord2018} and \ac{NLP}~\cite{nlp_Devlin2019,nlp_Lan2020,nlp_Lewis2020,nlp_Felbo2017,wang2020survey}. 
It has been introduced to \ac{SR} as well.
\ac{SSL} methods can enhance the representations of learned users/items and in a way to alleviate the data sparsity of cold-start users/items through exploring the intrinsic correlation among sequences of interactions~\cite{MIP2020,wu2020self,liu2021contrastive,yu2021self}.
Existing \ac{SSL} methods for \ac{SR} can be divided into two categories: \acfi{MIP} and \acfi{ICL}.
\ac{MIP} is inspired by masked language modeling in \ac{NLP}~\cite{nlp_Devlin2019}, which randomly masks items from the sequence and tries to predict the masked items based on the remaining information~\cite{ssl_mask_peterrec2020,MIP2020,bert4rec}.
Different from natural language, there is no natural grammatical constraint that a sequence of items satisfies.
Given the current sequence, \ac{ICL} samples negative target items to guide the model to learn the difference between the positive and negative target items~\cite{ssl_cl_queue2020,ssl_cl_s2s2020,ssl_cl_S3rec2020}.
The performance largely depends on the negative sampling strategy used~\cite{MoCo2020}.
These studies propose new \ac{SSL} signals by exploring a current interaction sequence itself but neglect the relation between different sequences.
Moreover, so far \ac{SSL} methods have not made a direct connection to user preference extraction and representation.

\begin{figure}[ht]
\centering
\setlength{\belowcaptionskip}{-0.2cm}
\includegraphics[clip,trim=1mm 3mm 3mm 3mm, width=0.65\columnwidth]{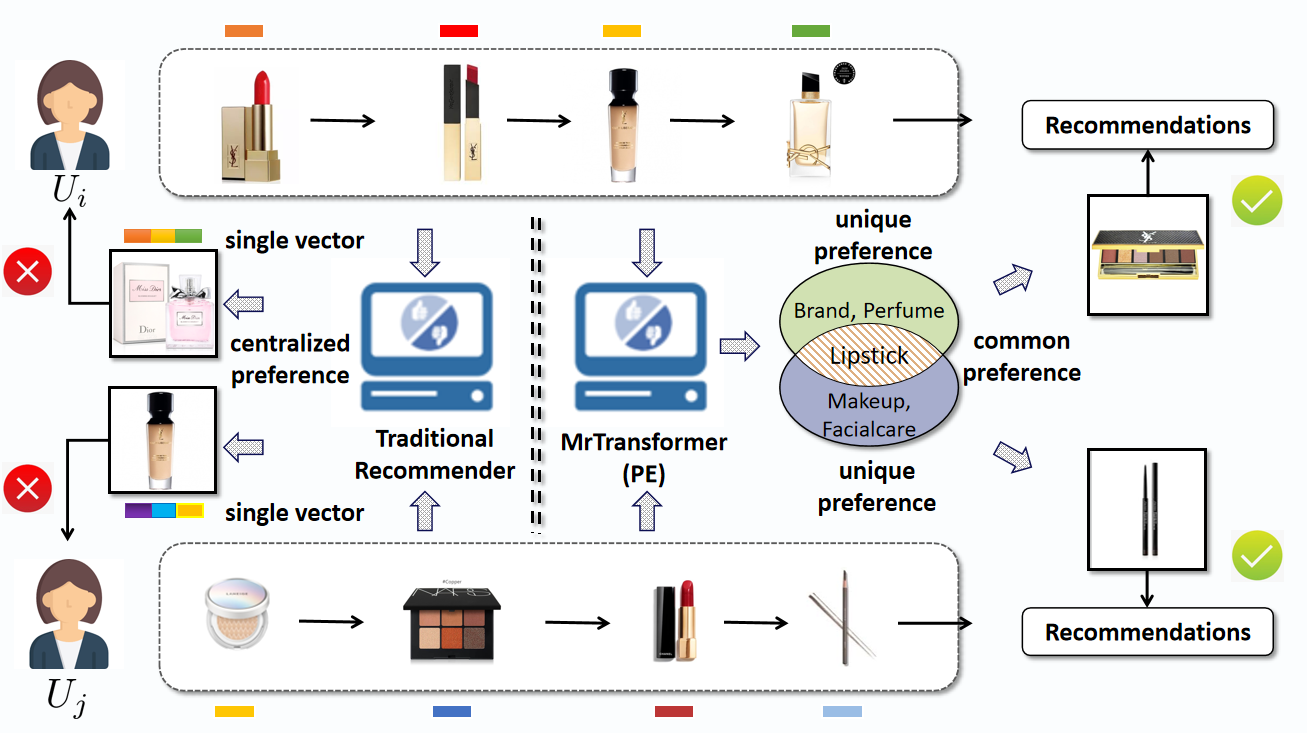}
\caption{Schematic comparison of a traditional recommender system and \OurMethod{} (PE).
Traditional recommender systems models the user preference as a single vector.
\OurMethod{} (PE) identifies common and unique preferences among different interaction sequences.}
\label{story}
\end{figure}

In this paper, we propose a novel learning strategy, named \textit{preference editing}, that focuses on discriminating the common and unique preference representations between different interaction sequences.
For example, in Figure~\ref{story}, users $U_i$ and $U_j$ have interacted in different ways with a recommender system in the beauty domain, giving rise to two interaction sequences.
Their common preference is lipstick. 
User $U_i$ also prefers perfume and this brand, and user $U_j$ prefers facial care products.
Our proposed sequential recommender \OurMethod{} (with \textbf{P}reference \textbf{E}diting) is able to identify common and unique preferences so as to get better preference representations.

Preference editing involves two operations to learn to edit the user preferences in a given pair of sequences: \textit{preference separation} and \textit{preference recombination}.
The former separates the common and unique preference representations for a pair of sequences.
The latter swaps the common user preference representations so as to get recombined user preference representations for each interaction sequence.
Specifically, given a pair of interaction sequences, we first obtain their preference representations $P_1$ and $P_2$. 
Then, we extract the common preference representations $c_1$ and $c_2$, and the unique preference representations $u_1$ and $u_2$, respectively, through the preference separation operation.
After that, we combine $c_2$ and $u_1$ to get the recombined preferences $P'_1$, and we combine $c_1$ and $u_2$ to get the recombined preferences $P'_2$ through the preference recombination operation.
Based on the above process, we can devise two types of \ac{SSL} signals:
\begin{enumerate*}
\item We use the recombined preferences to predict the next item in both sequences; and
\item We require that the recombined preferences (e.g., $P'_1$) are as similar as possible to the original preferences (e.g., $P_1$), and that the common preference representations are close to each other (i.e., $c_1=c_2$).
\end{enumerate*}
By doing so, we force the preference extraction model to learn how to identify and edit user preferences so as to do better user preference extraction and representation.

To extract multiple user preferences, we propose a transformer-based network, named \OurMethod{}.
Most studies on \ac{SR} assume that the user preference is centralized~\cite{ye2020time,Noninvasive,masr2020}, i.e., they assume that there is only one main user preference for each sequence.
Consequently, they usually model the user preference by a single vector.
For example, in Figure~\ref{story}, the traditional methods only obtain a mixed preference representation for each sequence while they ignore relations between other sequences.
Recent research has investigated how to identify multiple preferences through a multi-head attention mechanism on top of \acs{RNN}-based methods~\cite{diver_ijcai15,multi-pre-cikm20,guo2020session}.
But to the best of our knowledge, this has not been explored in transformer-based methods.
Therefore, we incorporate a \textit{preference identification} module into a transformer-based model, BERT4Rec~\cite{bert4rec}, which extracts multiple user preferences by concatenating some special tokens at the start of the each sequence.
When feeding an interaction sequence to the \OurMethod{}, we add some special tokens in the beginning with each capturing a particular of preference by attending to different items in the sequence.
The outputs corresponding to the special tokens are considered as preference representations.
To make sure that the preference representations differ from one another and capture different user preferences from the sequence, we devise an extra regularization terms.
Specifically, we maintain an attention vector for each preference so as to keep track of attention distributions for all items in the sequences being considered, which encourages coverage of the entire sequence.

To assess \OurMethod{} (PE), we carry out extensive experiments on two benchmark datasets: Amazon-Beauty and ML-100k.
The results show that \OurMethod{} with \textit{preference editing} significantly outperforms state-of-the-art baselines on both datasets in terms of Recall, MRR and NDCG.
We also find that long sequences whose user preferences are harder to extract and represent benefit the most from preference editing.
Especially, when the sequence length is between 20 and 30, preference editing achieves the biggest improvements.

To sum up, the main contributions of this work are as follows:
\begin{itemize}[leftmargin=*,nosep]
\item We devise a novel \acl{SSL} method, \textit{preference editing}, for \acp{SR}.
\item We propose a multi-preference transformer-based model, \OurMethod{}, for \acp{SR}.
\item We demonstrate the effectiveness of \OurMethod{} and \textit{preference editing} through extensive experiments on two benchmark datasets.
\end{itemize}


\section{Related Work}

In this section, we survey related work from two categories: trans\-former-based \ac{SR} and \acl{SSL} for \ac{SR}.

\subsection{Transformer-based \acl{SR}}
Various neural architectures or mechanisms have  successfully been applied to \acp{SR}.
These include \acp{RNN}~\cite{GRU4rec,NARM,ma2019pi,sun2018attentive}, \acp{CNN}~\cite{Caser2018,sun2020generic}, \acp{GNN}~\cite{wang2020global,qiu2020gag,yu2020tagnn}, \ac{RL}~\cite{xin2020self,wang2020kerl}, copy mechanism~\cite{repeatnet}, memory networks~\cite{KSR,chen2018sequential_memNet,wang2019collaborative}. 
Trans\-former-based methods have recently been proven to be effective ~\cite{SASRec,bert4rec,xie2020contrastive}.

\citet{SASRec} propose SASRec which introduces a self-attention mechanism (the most important component in the transformer) to \acp{SR} to identify important items from interactions.
Several variants have been proposed to improve upon SASRec.
\citet{mi2020ader} argue that traditional \acp{SR} use SASRec as a basic sequence representation extractor, and propose a continual learning setup with an adaptive distillation loss to update the recommender periodically as new data streams in.
\citet{li2020time} propose TiSASRec, a variant of SASRec, to explicitly model the timestamps of items in sequences to explore the influence of different time intervals on next item prediction.
\citet{luocollaborative} argue that even the same item can be represented differently for different users at the same time step; they propose a collaborative self-attention network to learn sequence representations and predict the preferences of the current sequence by investigating neighborhood sequences.
\citet{bert4rec} adopt a bidirectional transformer to predict masked items in sequences based on the surrounding items.
\citet{wang2020next} equip the transformer with hyper-graph neural networks to capture dynamic representations of items across time and users.
\citet{xie2020contrastive} propose three data augmentation approaches (crop/mask/reorder) to pre-train a transformer-based model to get user and sequence representations, and then fine-tune it on the \ac{SR} task.

Previous has also investigated how to combine auxiliary tasks or information with \ac{SR} based on transformer.
\citet{cho2020meantime} introduce multiple types of position embeddings by considering timestamp information, and propose a self-attention based model, in which each attention head uses a different position embedding.
\citet{wu2020deja} point out that previous studies ignore the temporal and context information when modeling the influence of a historical item to the current prediction; they propose a contextualized temporal attention mechanism to weigh the  influence of historical interactions not only on what items to interact with, but also when and how the interactions took place.
\citet{lin2020fissa} argue that modeling users' global preferences only based on their historical interactions is imperfect and the users' preference is uncertain; they propose a FISSA solution, which fuses item similarity models with self-attention networks to balance the local and global user preference representations by taking the information of the candidate items into account.
\citet{wu2020sse} argue propose a personalized transformer model with a recent regularization technique called \ac{SSE}~\cite{wu2019SSE} to overcome overfitting caused by simply adding user embeddings.

Although the studies listed above have proposed various trans\-former-based \ac{SR} models, they all assume that the user preference is centralized, i.e., they assume that there is a single main user preference for each sequence, and use a single vector to model the main user preference.
No previous work has considered how to identify users' multiple preferences behind the interaction sequence.
In contrast, we extract multiple user preferences and represent them using distributed vectors.

\subsection{Self-supervised learning for \ac{SR}}
Previous approaches to \ac{SR} are typically trained by predicting the next interaction, which is prone to suffer from data sparsity~\cite{MIP2020, xie2020contrastive}.
To mitigate this, some work explores \ac{SSL} and derives self-supervised signals to enhance the learning of item and sequence representations.

One category of \ac{SSL} approaches is \acf{MIP}, which randomly masks some items in an interaction sequence and tries to predict these masked items based on the remaining information.
\citet{bert4rec} adopt a BERT-like training scheme, which predicts the masked items in the sequence based on surrounding items.
Instead of masking items, \citet{MIP2020} propose to mask sparse categorical features of items; 
they mask or dropout some categorical feature embeddings to learn internal relations between two sets of categorical features.
\citet{ssl_mask_peterrec2020} represent and transfer user representation models for serving downstream tasks where only limited data exists through a pretrain-finetune strategy; in the pretraining stage, they randomly mask a certain percentage of items in the sequence and then predict the masked items to get user preference representations.

Another category of \ac{SSL} approaches to \ac{SR} is \acf{ICL}, which samples negative items to guide the model to learn the difference between positive and negative target items.
\citet{ssl_cl_queue2020} propose a fixed-size queue to store items' representations computed in previous batches, and use the queue to sample negative examples for each sequence.
\citet{ssl_cl_s2s2020} propose a sequence-to-sequence training strategy to mine extra supervision by looking at the longer-term future; they first use a disentangled encoder to obtain multiple representations of a given sequence and predict the representation of the future sub-sequence given the representation of the earlier sequence; sequence representations that are not from the same sequence as the earlier ones are considered as negative samples.
\citet{ssl_cl_S3rec2020} devise four auxiliary self-supervised objectives to learn correlations among four types of data (item attributes, items, sub-sequences, and sequences), respectively, by utilizing the mutual information maximization principle.
\citet{xie2020contrastive} propose three data augmentation methods (crop/mask/reorder); then they encode the sequence representation by maximizing the agreement between different augmented methods of the same sequence in the latent space.
\citet{xia2020self} model sequences as a hypergraph and propose a dual channel hypergraph convolutional network to capture higher-order relations among items within sequences; during training, they maximize the mutual information between the sequence representations learned via the two channels;
negative sampling \ac{SSL} ensures that different channels of the same sample are similar.

The \ac{SSL} strategies proposed in the studies listed above are mostly based on the current sequence itself.
In contrast, we focus on how to devise self-supervision signals by investigating the correlation of different sequences, i.e., forcing the \ac{SR} model to learn better representations by identifying the common and unique preferences for any pair of interaction sequences. 


\section{Method}

We first formulate the \ac{SR} task.
Then, we introduce our basic trans\-former-based multi-preference extraction model \OurMethod{}.
Next, we describe our new \ac{SSL} method \textit{preference editing}. 
Together, preference editing and \OurMethod{} constitute our complete model \OurMethod{} (PE).

\subsection{Task definition}
Let $\mathbb{I} = \{i_1, i_2, \ldots, i_{|\mathbb{I}|}\}$ denote the item set, and $\mathbb{S} = \{S_1, S_2, \ldots, S_{|\mathbb{S}|}\}$ denote the interaction sequence set.
Each interaction sequence $S \in \mathbb{S}$ can be denoted as $S = \{i_1, i_2, \ldots, i_\tau,\ldots, i_{t}\}$ where $i_\tau$ refers to the item interacted with at timestep $\tau$.
Given $S$, \ac{SR} aims to predict the next item that the user will interact with at timestep $t+1$ by computing the recommendation probabilities over all candidate items as follows:
\begin{equation}
\label{formalization}
\begin{split}
P(i_{t+1}|S) & \sim f(S), 
\end{split}
\end{equation}
where $P(i_{t+1}|S)$ denotes the probability of recommending the next item $i_{t+1}$, and $f(S)$ is the model or function to estimate $P(i_{t+1}|S)$.

\subsection{\OurMethod}
\label{section:ourmethod}
In this section, we will introduce our proposed transformer-based multi-preference extraction model \OurMethod{}.
Unlike the basic BERT4Rec model, the core idea of \OurMethod{} is to mine multiple preferences behind this interaction sequence, and then generate predictions based on the multiple preference representation.

\OurMethod{}'s network structure is based on the transformer, and it consists of three main components:
\begin{enumerate*}[label=(\arabic*)] 
\item a sequence encoder, 
\item a preference identification module, and 
\item a sequence decoder.
\end{enumerate*}
The preference identification module identifies multiple preference behind the current interaction sequence and represents it as distributed vectors.
Unlike RNN-based methods and traditional transformer-based methods, we concatenate $K$ special tokens ($\left[ P_1 \right], \\ \left[ P_2 \right], \ldots, \left[ P_K \right]$) at the start of each sequence, where each special token represents a particular user preference.
Next, we introduce these modules in detail.

\subsubsection*{\textbf{Sequence encoder}}
Here, we define the processed sequence $S' = \{ \left[ P_1 \right],$ $ \left[ P_2 \right], \ldots, \left[ P_K \right], i_1, i_2,$ $\ldots, i_\tau, \ldots, i_{t}\}$ which concatenates $K$ special tokens at the start of the sequence $S$.
It is worth noting that $K$ represents the number of latent preferences for the whole dataset, not for a particular sequence.
In this module, we encode the processed sequence $S^{'}$ into hidden representations.

First, we initialize the embedding matrix $\mathbf{E}$ of $S^{'}$, where $\mathbf{e_{i_{\tau}}} \in R^d$ represents the embedding for item $i_{\tau}$, $d$ is the embedding size.
Then we add the position embedding matrix $\mathbf{P}$ of $S^{'}$ it, which can be defined as $\mathbf{E} = \mathbf{E} + \mathbf{P}$.
After that, we feed the sequence of items into a stack of $L$ bidirectional transformer layers.
Each layer iteratively revises the representation of all positions by exchanging information across some specific positions, which are controlled by the masking matrix at previous layers.
This process is defined as follows:
\begin{equation}
\label{encoder_trm}
\begin{split}
\mathbf{E^{l}} = \operatorname{Trm}(\mathbf{E^{l-1}, Mask_e}), \\
\end{split}
\end{equation}
where $\operatorname{Trm}$ refers to the transformer layer, $\mathbf{E^{l}} \in R^{(K+t)*h}$ is the representation matrix of $S^{'}$ at the $l$-th layer, $h$ is the hidden size, and $\mathbf{Mask_e}$ is the masking matrix.
Specifically, each special token $\left[ P_k \right]$ can obtain information from all positions because it aims to capture the user's multiple preference behind the whole sequence.
So the receptive field is all positions, and the masking vector of each special token is all one.
For each item, the special tokens are masked out.
Therefore, masking vector of each item is composed of $K$ zeros and $t$ ones.
From the top layer, we can obtain the representation $\mathbf{E^L}$ of $S^{'}$. 

\subsubsection*{\textbf{Preference identification}}
In this module, we calculate the distributed attention scores over all items for each special token.
The process is defined as follows:
\begin{equation}
\label{identi_trm}
\begin{split}
\mathbf{P,A} = \operatorname{Ident}(\mathbf{E^{L}, Mask_I}), \\
\end{split}
\end{equation}
where $\operatorname{Ident}$ is implemented by a transformer layer, and $\mathbf{Mask_I}$ is the masking matrix for preference identification.
Different from $\mathbf{Mask_e}$, each special token only calculates attention weights over all items, so it masks out other special tokens.
And for each item, the masking vector is the same as $\mathbf{Mask_e}$.
$\mathbf{P} \in R^{K*h}$ is the representation matrix for the multiple preference corresponding to the first $K$ special tokens. 
We can also get the attention matrix $\mathbf{A}$ over all items for the special tokens.

To avoid that different special tokens focus on the same items and thus learn similar representations, we introduce a \emph{preference coverage mechanism}.
We maintain $K$ coverage vectors $c_{\tau}^k|_{k=1}^{K}$.
Vector $c_{\tau}^k$ is the sum of attention distributions over items at previous timesteps by the special token $\left[ P_k \right]$, which represents the degree of coverage that those items have received from the attention mechanism by $\left[ P_k \right]$ so far:
\begin{equation}
\label{coverage_vector}
\begin{split}
c_{\tau}^k = \sum_{j=1}^{\tau-1} a_{j}^k, \\
\end{split}
\end{equation}
where $a^k \in \mathbf{A}$ is a distributed attention vector over all items by $\left[ P_k \right]$ and $\sum_{i}a_i^k=1$.
And $c_{0}^k$ is a zero vector, which denotes that, at the first timestep, none of the items have been covered.

\subsubsection*{\textbf{Sequence decoder}}
After the preference identification module, we get the representation for the multiple preference.
Different from existing methods that use item representations to predict the next item~\cite{huang2019taxonomy,review_driven_sr2019,sentiSR2020,song2019session}, we use the learned multiple preference representation to do recommendation:
\begin{equation}
\label{decoder}
\begin{split}
P(i_{t+1}|S) = \operatorname{softmax}(\mathbf{p}W + b), \\
\end{split}
\end{equation}
where $\mathbf{p}$ is the sum of the multiple preference $\mathbf{P}$; $W$ is the embedding matrix of all items, and $b$ is the bias term.

\subsubsection*{\textbf{Objective functions}}
As with traditional \ac{SR} methods~\cite{hidasi2016parallel,wang2019collaborative,HRNN,STAMP}, our first objective is to predict the next item for each position in the input sequence.
We employ the negative log-likelihood loss to define the recommendation loss as follows:
\begin{equation}
\label{nll_loss}
\begin{split}
L_{rec}(\theta) & = -\frac{1}{|\mathbb{S}|} \sum_{S \in \mathbb{S}} \sum_{i_{\tau} \in S} \log P(i_{\tau+1}|S), \\
\end{split}
\end{equation}
where $\theta$ are all parameters of \OurMethod{}.

Apart from this, we also define a coverage loss to penalize repeatedly attending to the same items by different preferences:
\begin{equation}
\label{coverage_loss}
\begin{split}
L_{cov}(\theta) & = \sum_{k=1}^{K} \sum_{\tau} \min(a_{\tau}^k,c_{\tau}^k). \\
\end{split}
\end{equation}
Finally, the coverage loss function, weighted by the hyperparameter $\alpha$, is added to the recommendation loss function to yield the total loss function:
\begin{equation}
\label{total_loss}
\begin{split}
L(\theta) & = L_{rec}(\theta) + \alpha L_{cov}(\theta), \\
\end{split}
\end{equation}
where $\alpha$ controls the ratio of the coverage loss.

\begin{figure*}[ht]
\centering
\includegraphics[scale=0.45]{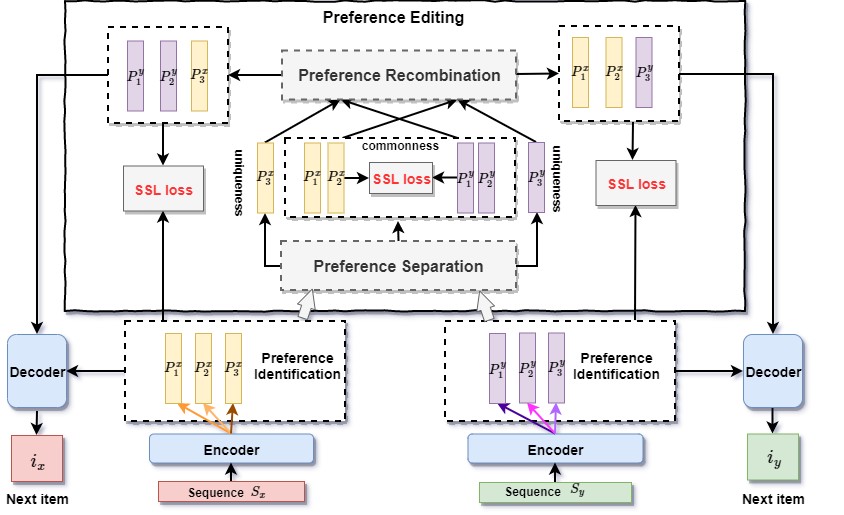}
\caption{Overview of \textit{preference editing}. Section~\ref{sec:pe} contains a walkthrough of this figure.}
\label{learning_figure}
\end{figure*}

\subsection{Preference editing}
\label{sec:pe}

In this subsection, we detail the \textit{preference editing} learning strategy to mine correlations among items between different interaction sequences.

As illustrated in Figure~\ref{learning_figure}, we first sample two sequences $S_x$ and $S_y$ from the sequence set $\mathbb{S}$ with the requirement that the two sequences share some common items while also have their unique items.
Through the \textit{preference identification} module, we get the multiple preference representation for each sequence (e.g., $P_1^x$, $P_2^x$ and $P_3^x$ for sequence $S_x$ and $P_1^y$, $P_2^y$ and $P_3^y$ for sequence $S_y$).
Then, the \textit{preference separation} module forces the model to separate common ($P_1^x$, $P_2^x$ and $P_1^y$, $P_2^y$) and unique ($P_3^x$ and $P_3^y$) preference representations for the paired sequences.
After that, the \textit{preference recombination} module swaps the common preference representation so as to obtain the recombined preference representations for each sequence (e.g., $P_1^y$, $P_2^y$, $P_3^x$ for sequence $S_x$ and $P_1^x$, $P_2^x$, $P_3^y$ for sequence $S_y$).
Next, we explain the above process in detail.

\subsubsection*{\textbf{Preference separation}}
For each pair of sequences $S_x$ and $S_y$, we get the multiple preference representation matrices $\mathbf{P_x}$ and $\mathbf{P_y}$ where $\mathbf{P_x}, \mathbf{P_y} \in R^{K*h}$.
In order to measure the degree of similarity between these two representations, we calculate the similarity matrix $\mathbf{I} \in R^{K*K}$ to consider their relation.
Each element $I_{ij} \in R$ is calculated as: 
    \begin{equation}
    \label{interaction_matrix}
    \begin{split}
    I_{ij} = W_I^{\mathrm{T}} \left[ p_i; p_j; p_i \odot p_j \right], \\
    \end{split}
    \end{equation}
where $p_i$ and $p_j$ are vectors in the multiple preference representation matrices $\mathbf{P_x}$ and $\mathbf{P_y}$, respectively;
$\odot$ is the element-wise calculation, and $W_I \in R^{3h*1}$ are the parameters.
Based on this similarity matrix, we calculate the attention matrices $\mathbf{A_x}$ and $\mathbf{B_y}$, which reflect the attention distribution of the preference representations of one sequence to the preference representation in another sequence as follows:
    \begin{equation}
    \begin{split}
    \label{attention_matrx}
    \mathbf{A_x} &{} = \operatorname{softmax}_{row}(\mathbf{I}) \\
    \mathbf{B_y} &{} = \operatorname{softmax}_{col}(\mathbf{I}), \\
    \end{split}
    \end{equation}
where $\operatorname{softmax}_{row}$ and $\operatorname{softmax}_{col}$ refer to performing a $\operatorname{softmax}$ calculation in rows and columns, respectively.
After that, the common and unique preference representation of each sequence are calculated as follows:
    \begin{equation}
    \begin{split}
    \label{c_and_u}
    \mathbf{C_x} &{} = \mathbf{B_y} \odot \mathbf{P_x}, 
    \quad \mathbf{U_x}  = (1-\mathbf{B_y}) \odot \mathbf{P_x} \\
    \mathbf{C_y} &{} = \mathbf{A_x} \odot \mathbf{P_y},  
    \quad \mathbf{U_y}  = (1-\mathbf{A_x}) \odot \mathbf{P_y}, \\
    \end{split}
    \end{equation}
where $C_x$ and $C_y$ refer to the common preference representation of each sequence respectively.
And $U_x$ and $U_y$ are the unique preference representation respectively.
Here, $A_x$ and $B_y$ play the role of the gate functions which filter the common and unique preference representation.

\subsubsection*{\textbf{Preference recombination}}
Through the preference separation module, we get the common and unique preference representation of the two input sequences $S_x$ and $S_y$.
Then, we swap the common preference representation to form a recombined representation of each sequence as follows:
    \begin{equation}
    \begin{split}
    \label{swap}
    \mathbf{P_x^{'}} &{} = \operatorname{combine}[\mathbf{C_y}; \mathbf{U_x}; \mathbf{C_y} \odot \mathbf{U_x}] \\
    \mathbf{P_y^{'}} &{} = \operatorname{combine}[\mathbf{C_x}; \mathbf{U_y}; \mathbf{C_x} \odot \mathbf{U_y}]. \\
    \end{split}
\end{equation}
Based on the recombined representation, we define two types of supervision signals for learning:
\begin{equation}
\label{total_singal}
\begin{split}
L_{SSL}(\theta) &{} =L_{pred}(\theta)+L_{reg}(\theta). \\
\end{split}
\end{equation}
$L_{pred}(\theta)$ is to do next item prediction based on the recombined representation $P_x^{'}$ and $P_y^{'}$:
\begin{equation}
\label{predict_signal}
\begin{split}
L_{pred}(\theta) & = L_{rec}^x + L_{rec}^y, \\
\end{split}
\end{equation}
where $ L_{rec}^x$ and $L_{rec}^y$ are negative log-likelihood loss, which are the same as calculation in Eq.~\ref{nll_loss}.
And $ L_{rec}^x$ and $L_{rec}^y$ are used to predict the next item of each sequence using the recombined representation.


%
$L_{reg}(\theta)$ is a regularization term, which consists of three parts:
\begin{equation}
\label{predict_signal}
\begin{split}
L_{reg}(\theta)^x &{} = \frac{1}{K*h} \sum_{S_x, S_y \in \mathbb{S}}(\mathbf{P_x} - \mathbf{P_x^{'}})^2 \\
L_{reg}(\theta)^y &{} = \frac{1}{K*h} \sum_{S_x, S_y \in \mathbb{S}}(\mathbf{P_y} - \mathbf{P_y^{'}})^2 \\
L_{reg}(\theta)^c &{} = \frac{1}{K*h} \sum_{S_x, S_y \in \mathbb{S}}(\mathbf{C_x} - \mathbf{C_y})^2 \\
L_{reg}(\theta) &{} =L_{reg}(\theta)^x+L_{reg}(\theta)^y+L_{reg}(\theta)^c. \\
\end{split}
\end{equation}
$L_{reg}(\theta)^x$ and $L_{reg}(\theta)^y$ make sure that the recombined representation is close enough to the original preference representation, respectively.
$L_{reg}(\theta)^c$ requires that the learned common representations are also close to each other.

The final training loss for \OurMethod{} is as follows:
\begin{equation}
\label{model_learn_loss}
\begin{split}
L_{all}(\theta) &{} =L(\theta)+L_{SSL}(\theta), \\
\end{split}
\end{equation}
where $L(\theta)$ is the loss for preference extraction and recommendation (Eq.~\ref{total_loss}), $L_{SSL}(\theta)$ is the loss for preference editing (Eq.~\ref{total_singal}).

We use \textbf{\OurMethod{}} to refer to our basic model that models a users' preferences based on transformer layers by using the preference identification module as detailed in Section~\ref{section:ourmethod}.
We write \textbf{\OurMethod{} (PE)} for \OurMethod{} pretrained with the preference editing learning strategy described in this section.


\section{Experimental Setup}

We seek to answer the following questions in our experiments:
\begin{enumerate}[label=(RQ\arabic*)]
\item What is the performance of \OurMethod{} (PE) compared to other methods? Does it outperform the state-of-the-art methods in terms of Recall, MRR and NDCG on all datasets? 
\item What is the effect of the preference editing learning strategy on the performance of \OurMethod{} (PE)? And how does it affect sequences of different lengths?
\item What is the effect of the preference coverage mechanism on the performance of \OurMethod{} (PE)?
\item How does the hyperparameter $K$ (the number of assumed preferences) affect the performance of \OurMethod{} (PE)? 
\end{enumerate}

\subsection{Datasets}
We conduct experiments on five datasets in different domains:
\begin{itemize}[leftmargin=*,nosep]
\item \textbf{Amazon Beauty, Sports, Toys}: These datasets are obtained from Amazon product review datasets crawled by \citet{mcauley2015image}. 
We select three subcategories of Beauty (40,226 users and 54,542 items), Sports (25,598 users and 18,357 items) and Toys (19,412 users and 11,924 items).\footnote{\url{http://jmcauley.ucsd.edu/data/amazon/}}
\item \textbf{MovieLens}: This is a popular benchmark dataset for recommendation evaluation. We adopt a well-established version, ML-100k (943 users and 1,349 items).\footnote{\url{https://grouplens.org/datasets/movielens/}}
\item \textbf{Yelp}: This is a popular dataset for business recommendation (30,431 users and 20,033 items). \footnote{\url{https://www.yelp.com/dataset}}
As it is very large, we only use the transaction records after January
1st, 2019 like ~\cite{ssl_cl_S3rec2020}. 
\end{itemize}

\noindent%
For data processing, we follow~\cite{bert4rec,SASRec}.
We convert the rating scores of reviews to the implicit feedback of 1 if the user has interacted with this item.
Then we group the interactions by the same user and sort them according to the timesteps.
Like~\cite{zhang2019feature,review_driven_sr2019,meng2020incorporating}, since we do not target cold-start recommendation, we filter out cold-start users who have less than 5 interactions and cold-start items with less than 5 interactions.
Because each user has a lot of interactions in a long time, we limit the maximum length of each sequence to be 50.
And the statistics of datasets are shown in Table~\ref{dataset}.

\begin{table}[t]
\caption{Statistics of datasets.}
\label{dataset}
\setlength{\tabcolsep}{4pt}
\begin{tabular}{lrrrrrr}
\toprule
Datasets & \#users & \#items & \#actions & \#min(len) & \#max(len) & \#avg(len) \\
\midrule
Beauty    &   40,226  &  54,542 &  353,962   & 3 & 291 &  8.8   \\
Sports    &   25,598  &  18,357 &  296,337   & 3 & 294 &  6.3   \\
Toys      &   19,412  &  11,924 &  167,597   & 3 & 548 &  6.6   \\
ML-100k   &   943    &   1,349    &   99,287 & 17 & 646 & 103.2  \\
Yelp      &   30,431  &  20,033 &  316,354   & 3 & 348 &  8.4   \\
\bottomrule
\end{tabular}
\squeeze
\end{table}

\subsection{Evaluation metrics}

We adopt the leave-one-out recommendation evaluation scheme~\cite{bert4rec}.
We use the last item and the second last item as the test set and validation set, and the rest as training set.
We concatenate the training and validation data to predict the last item in the test set.
Following~\cite{review_driven_sr2019,mcauley2015image}, we pair each ground truth item with 100 randomly sampled negative items according to their popularities.

We report results using Recall@k, MRR@k and NDCG@k with $k=5, 10, 20$.
Assuming that the tested item of user $u$ is ranked $r_u$ based on the predicted scores, the metrics are calculated as follows:
\begin{itemize}[leftmargin=*,nosep]
\item Recall@k: The proportion of cases when the ground-truth item is amongst the top-$k$ ranked items.
\item MRR@k: The average of reciprocal ranks of the ground-truth items, i.e., $MRR_u@k = \frac{1}{r_u}$ if $r_u \le k$ and $MRR_u@k = 0$ otherwise.
\item NDCG@k: A position-aware metric which assigns larger weights on higher ranks, i.e., $NDCG_u@k = \frac{1}{\log_2(1+r_u)}$ if $r_u \le k$ and $NDCG_u@k = 0$ otherwise.
\end{itemize}

\begin{table*}[ht]

\centering
\caption{Experimental results on five datasets.
\textbf{Bold face} indicates the best results in terms of the corresponding metrics.
Significant improvements over the best baseline results are marked with $^*$ (t-test, $p < .05$).
}
\footnotesize
\label{results_1}
\setlength{\tabcolsep}{0.5mm}{
\begin{tabular}{llccccccccccc}

\toprule
 & Metric & POP & BPR-MF & Item-Knn & SASRec & SSE-PT & FISSA & BERT4Rec & S3-Rec & \OurMethod{} & \OurMethod{} (PE) \\
\midrule
\multirow{9}{*}{\rotatebox{90}{\bf Beauty}} & Recall@5 &  0.0315 & 0.1390 & 0.1660 & 0.1934 & 0.1959 & 0.2116 & 0.2137 & 0.2173 & 0.2220 & \textbf{0.2286}     \\
& Recall@10 & 0.0628 & 0.2152 & 0.2029 & 0.2653 & 0.2813 & 0.3079 & 0.2964 & 0.2909 & 0.2989 & \textbf{0.3126} \\
& Recall@20 & 0.1280 & 0.3292 & 0.2695 & 0.3724 & 0.3901 & 0.4289 & 0.3971 & 0.3926 & 0.4070 & \textbf{0.4302}\rlap{$^*$}\\
& MRR@5 & 0.0147 & 0.0755 & 0.1259 & 0.1112 &  0.1199 & 0.1198 & 0.1299 & 0.1348 & 0.1438 & \textbf{0.1444}\rlap{$^*$} \\
& MRR@10 & 0.0187 & 0.0855 & 0.1308 & 0.1220 & 0.1286 & 0.1325 & 0.1408 & 0.1445 & 0.1540 & \textbf{0.1555}\rlap{$^*$} \\
& MRR@20 & 0.0230 & 0.0933 & 0.1352 & 0.1291 & 0.1327 & 0.1408 & 0.1477 & 0.1514 & 0.1614 & \textbf{0.1635}\rlap{$^*$} \\
& NDCG@5 & 0.0188 & 0.0912 & 0.1359 & 0.1436 & 0.1491 & 0.1424 & 0.1506 & 0.1552 & 0.1632 & \textbf{0.1653}  \\
& NDCG@10 & 0.0288 & 0.1157 & 0.1477 & 0.1633 &  0.1683 & 0.1734 & 0.1773 & 0.1789 & 0.1879 & \textbf{0.1923}\rlap{$^*$}\\
& NDCG@20 & 0.0449 & 0.1443 & 0.1644 & 0.1823 & 0.1944 & 0.2039 & 0.2026 & 0.2044 & 0.2151 & \textbf{0.2218}\rlap{$^*$}\\
\midrule
\multirow{9}{*}{\rotatebox{90}{\bf Sports}} & Recall@5 & 0.0313 & 0.0376 & 0.1512  & 0.1622 & 0.2102 &	0.2133 & 0.2124 & 0.2072 &	0.2069 & \textbf{0.2320}\rlap{$^*$} \\
& Recall@10 & 0.0624 & 0.0819 & 0.1927  & 0.2788 & 0.3212 & 0.3180 & 0.3105 & 0.3250 & 0.3225	& \textbf{0.3456}\rlap{$^*$}    \\
& Recall@20 & 0.1364 & 0.1735 & 0.2260  & 0.3908 & 0.4492 & 0.4592 & 0.4464	& 0.4817 & 0.4829 & \textbf{0.5000}\rlap{$^*$}     \\
& MRR@5 & 0.0139 & 0.0167 & 0.0449 & 0.0913 & 0.1180 & 0.1262 & 0.1227 &0.1122 & 0.1146 &\textbf{0.1354}     \\
& MRR@10 & 0.0178 & 0.0224 & 0.0547  & 0.1042 & 0.1344 & 0.1367 & 0.1356 & 0.1276 & 0.1297 & \textbf{0.1504}\rlap{$^*$}     \\
& MRR@20 & 0.0227 & 0.0285 & 0.1224  & 0.1105 & 0.1464 & 0.1486 & 0.1449 & 0.1384 & 0.1408 & \textbf{0.1616}\rlap{$^*$}     \\
& NDCG@5 & 0.0181 & 0.0218 & 0.0783  & 0.1102 & 0.1407 & 0.1325 & 0.1449 & 0.1356 & 0.1374 & \textbf{0.1593}\rlap{$^*$}     \\ 
& NDCG@10 & 0.0279 & 0.0359 & 0.1132 & 0.1532 & 0.1780 & 0.1726 & 0.1764 & 0.1734 & 0.1744 & \textbf{0.1958}\rlap{$^*$}    \\ 
& NDCG@20 & 0.0464 & 0.0587 & 0.1550  & 0.1704 & 0.2126 & 0.2157 & 0.2106 & 0.2128 & 0.2149 & \textbf{0.2347}\rlap{$^*$}    \\ 
\midrule
\multirow{9}{*}{\rotatebox{90}{\bf Toys}} & Recall@5 & 0.0348 & 0.0487 & 0.0986  & 0.2411 & 0.2940 & 0.3127 & 0.2980 & 0.3110 & 0.3023 & \textbf{0.3285}\rlap{$^*$} \\
& Recall@10 & 0.0697 & 0.0970 & 0.2344  & 0.3417 & 0.4001 & 0.4062 & 0.3878 & 0.4094 & 0.3955 &  \textbf{0.4218}    \\
& Recall@20 & 0.1443 & 0.1995 & 0.2360  & 0.4407 & 0.5344 & 0.5134 & 0.5058 & 0.5282 & 0.5224 &  \textbf{0.5445}\rlap{$^*$}     \\
& MRR@5 & 0.0161 & 0.0226 & 0.0388  & 0.1530 & 0.1753 & 0.1934 & 0.1932 & 0.1964 & 0.1969 & \textbf{0.2208}     \\
& MRR@10 & 0.0206 & 0.0288 & 0.0562  & 0.1630 & 0.1882 & 0.2098 & 0.2050 & 0.2096 & 0.2092 & \textbf{0.2332}\rlap{$^*$}     \\
& MRR@20 & 0.0256 & 0.0356 & 0.0665  & 0.1707 & 0.1960 & 0.2200 & 0.2131 & 0.2177 & 0.2178 & \textbf{0.2415}\rlap{$^*$}     \\
& NDCG@5 & 0.0207 & 0.0290 & 0.0534  & 0.1749 & 0.2047 & 0.2254 & 0.2192 & 0.2249 & 0.2231 &\textbf{0.2476}\rlap{$^*$}     \\ 
& NDCG@10 & 0.0318 & 0.0443 & 0.0965 & 0.2177 & 0.2380 & 0.2653 & 0.2481 & 0.2567 & 0.2530 & \textbf{0.2776}\rlap{$^*$}    \\ 
& NDCG@20 & 0.0505 & 0.0699 & 0.1325  & 0.2274 & 0.2708 & 0.3025 & 0.2778 & 0.2866 & 0.2849 &\textbf{0.3084}\rlap{$^*$}    \\ 
\midrule
\multirow{9}{*}{\rotatebox{90}{\bf ML-100k}} & Recall@5 & 0.0965 & 0.1866 & 0.1845 & 0.2948 & 0.3017 & 0.2585 & 0.3003 & 0.3012 & 0.3436 & \textbf{0.3601}\rlap{$^*$} \\
& Recall@10 & 0.1431 & 0.3138 & 0.3276 & 0.4746 & 0.4688 & 0.4492& 0.4662 & 0.4761 & 0.4920 &  \textbf{0.5005}    \\
& Recall@20 & 0.2396 & 0.4655 & 0.4835 & 0.6548 & 0.6723 & 0.6409& 0.6506 & 0.6755 & 0.6648&  \textbf{0.6808}\rlap{$^*$}     \\
& MRR@5 & 0.0484 & 0.0989 & 0.0958 & 0.1753 & 0.1748 & 0.1311 & 0.1707 &0.1513 & 0.1929 &\textbf{0.2031}     \\
& MRR@10 & 0.0549 & 0.1154 & 0.1152 & 0.1984 & 0.2046 & 0.1555 & 0.1926 & 0.1748 & 0.2119 & \textbf{0.2211}\rlap{$^*$}     \\
& MRR@20 & 0.0613 & 0.1258 & 0.1257 & 0.2117 & 0.2183 & 0.1690 & 0.2058 & 0.1897 & 0.2230 & \textbf{0.2336}\rlap{$^*$}     \\
& NDCG@5 & 0.0602 & 0.1203 & 0.1176 & 0.2035 & 0.2106 & 0.1624 & 0.2028 & 0.1881 & 0.2315 &\textbf{0.2416}\rlap{$^*$}     \\
& NDCG@10 & 0.0755 & 0.1610 & 0.1642 & 0.2655 & 0.2599 & 0.2230 & 0.2562 & 0.2448 & 0.2784 & \textbf{0.2866}\rlap{$^*$}    \\
& NDCG@20 & 0.0996 & 0.1992 & 0.2032 & 0.3088 & 0.2993 & 0.2718 & 0.3033 & 0.2985& 0.3209 &\textbf{0.3321}\rlap{$^*$}    \\
\midrule
\multirow{9}{*}{\rotatebox{90}{\bf Yelp}} & Recall@5 & 0.0474 & 0.0500 & 0.0499  & 0.4070 & 0.3884 & 0.4520 & 0.4561 & 0.4224 & 0.4403 & \textbf{0.4640}\rlap{$^*$} \\
& Recall@10 & 0.0960 & 0.0983 & 0.1080  & 0.5691 & 0.5609 & 0.6153 & 0.6138 & 0.5926 & 0.6086 &  \textbf{0.6312}    \\
& Recall@20 & 0.1897 & 0.1970 & 0.2278  & 0.7414 & 0.7684 & 0.7757 & 0.7621 & 0.7625 & 0.7771 &  \textbf{0.7972}\rlap{$^*$}     \\
& MRR@5 & 0.0227 & 0.0230 & 0.0224  & 0.2265 & 0.2261 & 0.2690 & 0.2703 & 0.2439 & 0.2609 &\textbf{0.2742}     \\
& MRR@10 & 0.0290 & 0.0307 & 0.0298  & 0.2476 & 0.2492 & \textbf{0.2974} & 0.2914 & 0.2666 & 0.2834 & 0.2965     \\
& MRR@20 & 0.0353 & 0.0365 & 0.0380  & 0.2592 & 0.2686 & 0.3002 & 0.3017 & 0.2783 & 0.2951 & \textbf{0.3080}\rlap{$^*$}     \\
& NDCG@5 & 0.0288 & 0.0291 & 0.0291  & 0.2699 & 0.2663 & 0.3140 & 0.3163 & 0.2880 & 0.3053 &\textbf{0.3211}\rlap{$^*$}     \\ 
& NDCG@10 & 0.0443 & 0.0447 & 0.0476  & 0.3242 & 0.3224 & 0.3631 & 0.3673 & 0.3430 & 0.3597 & \textbf{0.3752}\rlap{$^*$}    \\ 
& NDCG@20 & 0.0677 & 0.0691 & 0.0777  & 0.3632 & 0.3789 & 0.4089 & 0.4048 & 0.3860 & 0.4023 &\textbf{0.4172}\rlap{$^*$}    \\ 
\bottomrule

\end{tabular}
}
\end{table*}

\subsection{Methods used for comparison}
To assess the effectiveness of \OurMethod{} and \OurMethod{} (PE), we compare it with the following methods:
\begin{itemize}[leftmargin=*,nosep]
\item \textbf{POP} ranks items in the training set based on their popularities, and always recommends the most popular items~\cite{NCF}.

\item \textbf{BPR-MF} is a commonly used matrix factorization method.
We apply it for \ac{SR} by representing a new sequence with the average latent factors of items appearing in the sequence so far~\citep{GRU4rec}.

\item \textbf{Item-KNN} computes an item-to-item similarity matrix and recommends items that are similar to the actual item.
Regularization is included to avoid coincidental high similarities~\cite{itemknn}.

\item \textbf{SASRec}~\cite{SASRec} uses a left-to-right self-attention based model to capture users’ sequential behavior.

\item \textbf{SSE-PT}~\cite{wu2020sse} is a personalized transformer-based model with a regularization technique called \ac{SSE}~\cite{wu2019SSE} to overcome overfitting caused by simply adding user embeddings.

\item \textbf{FISSA}~\cite{lin2020fissa} uses the SASRec model to get the user preference representations, 
with a transformer-based layer that fuses item similarity to model the global user preference representation;
a gating module balances the local and global representations.

\item \textbf{BERT4Rec}~\cite{bert4rec} models the user preference representation with a bidirectional transformer network; during training, it randomly masks some items in the sequence and predicts these items jointly conditioned on their left and right context; during testing, it only masks the last item to do recommendations.

\item \textbf{S3-rec}~\cite{ssl_cl_S3rec2020} adopts SASRec as the base model and devises four auxiliary self-supervised objectives to pretrain the sequential recommender model using attribute, item, subsequence, and sequence by utilizing mutual information maximization; it fine-tunes the parameters according to the \ac{SR} task; we do not consider attribute characteristics and only use the \ac{MIP} loss function as they are not supported by the datasets.

\end{itemize}

\noindent%
Other \ac{SSL} based methods
\cite[e.g., ][]{ssl_mask_peterrec2020,MIP2020,ssl_cl_queue2020,wu2020self,yu2021self} have only been proposed for general, cross-domain or social recommendation tasks; hence, we omit comparisons against them.
We also exclude the \ac{SSL} based method in \cite{ssl_cl_s2s2020}, as the authors found flaws in their experimental setup.
As for the \ac{SSL} based method in \cite{xia2020self}, they adopt a different experimental setting of next-session recommendation to model the sequence graph representation, instead, we use the next-item recommendation setting which cannot meet their premise of hypergraph construction.

\subsection{Implementation details}
We implement POP, BPR-MF and Item-KNN using Tensorflow.
For SASRec,\footnote{\url{https://github.com/kang205/SASRec}} SSE-PT,\footnote{\url{https://github.com/wuliwei9278/SSE-PT}} FISSA,\footnote{\url{https://github.com/RUCAIBox/CIKM2020-S3Rec}} BERT4Rec,\footnote{\url{https://github.com/FeiSun/BERT4Rec}} and S3-rec,\footnote{\url{https://github.com/RUCAIBox/CIKM2020-S3Rec}} we use the code provided by the authors. We use hyperparameters as reported or suggested in the original papers.

We implement \OurMethod{} with Tensorflow.
All parameters are initialized using a truncated normal distribution in the range $\left[ -0.02, 0.02\right]$.
We set the hidden size to 64 and drop ratio to 0.5.
We set the number of transformer layers to 2 and the number of attention heads to 2.
For \OurMethod{} (PE), the training phase contains two stages (i.e., pre-training and fine-tuning stage).
The learned parameters in the pre-training stage are used to initialize the embedding layers and transformer layers in the fine-tuning stage.
In the pre-training stage, we only use the \ac{SSL} loss (as defined in Eq.~\ref{total_singal}) to train \OurMethod{}.
In the fine-tune stage, we use the recommendation loss (as defined in Eq.~\ref{predict_signal}) to train \OurMethod{}.
We try different settings for the preference size $K$, the analysis of which can be found in Section~\ref{analysis_K}.
We train the model using Adam optimizer, we set the learning rate as 1e-4, $\beta_1 = 0.99, \beta_2=0.999$, $l_2$ weight decay of 0.01, with linear decay of the learning rate. 
We also apply gradient clipping with range $\left[-5, 5\right]$ during training. 
To speed up training and converge, we use a mini-batch size of 256. 
We test the model performance on the validation set for every epoch.
All the models are trained on a GeForce GTX TitanX GPU.

For sampling sequence pairs in the learning of preference editing, we sample 20 sequences for each sequence according to its similarities with all other sequences.
Specifically, we first build a 0-1 vector for each sequence.
If the item appears in the sequence, the corresponding element is 1, otherwise it is 0.
Then we use the FAISS library\footnote{\url{https://github.com/facebookresearch/faiss}} to search the nearest neighbors for each sequence.
We choose the 20 samples according to their item frequency similarity, and make sure that each pair of sequences shares some common items but also have some unique items.


\section{Results and Analysis}

\subsection{Overall performance of \OurMethod{} (PE)}
\label{rq1}
To answer RQ1, we report on the performance of \OurMethod{} (PE) and the baseline methods in terms of Recall, MRR, NDCG. See Table~\ref{results_1}.
We obtain the following insights from the results.

First, \OurMethod{} (PE) achieves the best results on all datasets in terms of all metrics.
It outperforms strong Transformer-based \ac{SR} methods (SSE-PT, FISSA) and \ac{SSL}-based methods (BERT4Rec and S3-Rec).
Specifically, concerning the transformer-based \ac{SR} methods on the ``Beauty'' dataset, 
the increase over FISSA is 8.03\%, 20.53\% and 16.08\% in terms of Recall@5, MRR@5, and NDCG@5.
As for the \ac{SSL}-based \ac{SR} methods, 
the increase over S3-Rec is 9.58\%, 7.99\% and 8.51\% in terms of Recall@5, MRR@5, and NDCG@5.
On the ``ML-100k'' dataset, the increase over SSE-PT is 19.35\%, 15.96\% and 14.71\% in terms of Recall@5, MRR@5, and NDCG@5,
while the increase over S3-Rec is 19.55\%, 33.97\% and 28.44\% in terms of Recall@5, MRR@5, and NDCG@5.
On the ``Yelp'' dataset, the increase over the transformer-based \ac{SR} method FISSA is 2.77\%, 2.59\% and 3.33\% in terms of Recall@20, MRR@20, and NDCG@10.
And the increase over \ac{SSL}-based \ac{SR} method BERT4Rec is 4.60\%, 2.08\%and 3.06\% in terms of Recall@20, MRR@20, and NDCG@20.
To sum up, our method consistently outperforms all the compared methods on these datasets.

The improvements are mainly because that \OurMethod{} (PE) achieves better preference extraction and representation by discriminating the common and unique items of multiple preferences between sequences with the help of preference editing.
We will analyze the preference editing learning strategy in more depth in Section~\ref{analysis_pe}.

Second, the improvements of \OurMethod{} (PE) on ``ML-100k'' are larger than those on other datasets.
This is related to the characteristics of different datasets.
For example, the ``Beauty'' dataset is collected from cosmetic purchases.
For a particular sequence, it usually has a clear preference and specific buying needs, although the performance on the whole dataset can be very extensive and diverse.
This is also true for the ``Sports'', ``Toys'' and ``Yelp'' datasets, which are collected from the outdoor sports, game purchase records and business reviews, respectively.
Under these scenarios, users' preferences are usually fixed and clear within the same sequence.
However, the ``ML-100k'' dataset is collected from movie watching.
Users' watching preferences usually change from time to time with the change of context such as time, place or mood, even within the same sequence.
Since \OurMethod{} (PE) is especially adept at extracting multiple preferences and capturing their commonness and uniqueness through preference editing, its advantages are more obvious on ``ML-100k''.

Third, the improvements of NDCG and MRR on most datasets are larger than Recall generally.
On the one hand, this demonstrates that \OurMethod{} (PE) is beneficial to the ranking of the recommendation list.
On the other hand, this also reveals that Recall is generally more difficult to improve.
Compared with the improvements of the best baseline over the second best, i.e., the largest increase of S3-Rec over FISSA on the ``Beauty'' dataset achieves 2.69\% in terms of Recall@5.
On the ``ML-100k'' dataset, the largest increase of SSE-PT over BERT4Rec achieves the improvement of 3.33\% in terms of Recall@20, the improvements of \OurMethod{} (PE) on Recall is already considered large.
Generally, \ac{SSL}-based \ac{SR} methods outperform transformer-based \ac{SR} methods, which indicates that \ac{SSL} is an effective direction for \acp{SR} by introducing more supervision signals to improve representation learning.
What's more, the \ac{SSL}-based method S3-Rec is better than the other \ac{SSL}-based methods, e.g., BERT4Rec, on almost all datasets.
This is because S3-Rec also leverages sequence-level self-supervised signals besides the item-level signals, which further confirms that the self-supervised signals are very helpful for improving the recommendation performance.

\subsection{Effect of preference editing}
\label{analysis_pe}
To answer RQ2, we conduct an ablation study to analyze the effects of the preference editing learning strategy.
We compare \OurMethod{} (PE) with \OurMethod{} where \OurMethod{} is only trained with the recommendation loss (Eq.~\ref{nll_loss}) and the preference coverage loss (Eq.~\ref{coverage_loss}).
The results are shown in Table~\ref{results_1}.

The results decrease when removing preference editing on all datasets.
By comparison, the results drop by more than 5.70\%pt, 1.30\%pt and 3.11\%pt in terms of Recall@20, MRR@20 and NDCG@20 on the ``Beauty'' dataset.
And the results drop by more than 4.80\%pt, 5.08\%pt and 4.36\%pt in terms of Recall@5, MRR@5 and MRR@5 on the ``ML-100k'' dataset.
On the ``Yelp'' dataset, the results drop by more than 2.58\%pt, 4.37\%pt and 3.70\%pt in terms of Recall@20, MRR@20 and NDCG@20, respectively.
This demonstrates that preference editing is effective by forcing \OurMethod{} to learn common and unique items between sequences.
At the same time, we see that the MRR and NDCG of \OurMethod{} are slightly higher than the strong baselines like BERT4Rec and FISSA while the Recall values are comparable on most datasets.
This is reasonable because \OurMethod{} is also a transformer-base method, and without preference editing, its improvements are limited.

We also carry out analysis experiments on the three datasets (i.e., ``Sports'', ``ML-100k'', ``Yelp'') to analyze the effects of preference editing on sequences with different lengths.
Specifically, we divide the sequences into four groups according to their length.
The results are shown in Table~\ref{analysis_PE_sports1}--\ref{analysis_PE_yelp2}.

\begin{table}[ht]
\centering
\caption{Recall, MRR and NDCG of \OurMethod{} without preference editing on the ``Sports'' dataset. The best results are denoted in underline fonts.}
\label{analysis_PE_sports1}
\setlength{\tabcolsep}{1.5mm}{
\begin{tabular}{lccccccccc}
\toprule
\multirow{3}{*}{Length} & \multicolumn{9}{c}{\bf \OurMethod{}} \\
\cmidrule(r){2-10}
& \multicolumn{3}{c}{Recall} & \multicolumn{3}{c}{MRR} & \multicolumn{3}{c}{NDCG} \\
\cmidrule(r){2-4}
\cmidrule(r){5-7}
\cmidrule(r){8-10}
& @5 & @10 & @20 & @5 & @10 & @20 & @5 & @10 & @20 \\
\midrule 
$\left[ 20,30 \right)$ & \underline{0.2176} & \underline{0.3387} & \underline{0.4846} & 0.1267 & 0.1424 & 0.1525 & 0.1492 & \underline{0.1879} & \underline{0.2246}  \\
$\left[ 30,40 \right)$ & 0.2172& 0.3252 & 0.4521 & \underline{0.1317} & \underline{0.1455} & \underline{0.1543} & \underline{0.1528} & 0.1871 & 0.2192 \\
$\left[ 40,50 \right]$ & 0.2018 & 0.3211 & 0.4678 & 0.1233 & 0.1392 & 0.1497 & 0.1425 & 0.1810 & 0.2185 \\
$\geq 50$ & 0.2125 & 0.3375 & 0.4560 & 0.1295 & 0.1430 & 0.1520 & 0.1486 & 0.1822 & 0.2153  \\
\bottomrule
\end{tabular}}
\end{table}

\begin{table}[ht]
\centering
\caption{Recall, MRR and NDCG of \OurMethod{} with preference editing on the ``Sports'' dataset. The best results are denoted in underline fonts.}
\label{analysis_PE_sports2}
\setlength{\tabcolsep}{1.5mm}{
\begin{tabular}{lccccccccc}
\toprule
\multirow{3}{*}{Length} & \multicolumn{9}{c}{\bf \OurMethod{} (PE)} \\
\cmidrule(r){2-10}
& \multicolumn{3}{c}{Recall} & \multicolumn{3}{c}{MRR} & \multicolumn{3}{c}{NDCG} \\
\cmidrule(r){2-4}
\cmidrule(r){5-7}
\cmidrule(r){8-10}
& @5 & @10 & @20 & @5 & @10 & @20 & @5 & @10 & @20 \\
\midrule 
$\left[ 20,30 \right)$ & 0.2263 & 0.3413 & \underline{0.4963} & 0.1347 & 0.1495 & 0.1602 & 0.1573 & 0.1940 & 0.2331  \\
$\left[ 30,40 \right)$ & \underline{0.2496} & \underline{0.3589} & 0.4958 & 0.1445 & 0.1586 & 0.1683 & 0.1705 & \underline{0.2089} & \underline{0.2457} \\
$\left[ 40,50 \right]$ & 0.2247 & 0.3394 & 0.4908 & 0.1376 & 0.1523 & 0.1625 & 0.1591 & 0.1956 & 0.2334 \\
$\geq 50$ & 0.2313 & 0.3438 & 0.4938 & \underline{0.1523} & \underline{0.1665} & \underline{0.1770} & \underline{0.1718} & 0.2073 & 0.2403  \\
\bottomrule
\end{tabular}}
\end{table}

\begin{table}[ht]
\centering
\caption{Recall, MRR and NDCG of \OurMethod{} without preference editing on the ``ML-100k'' dataset. The best results are denoted in underline fonts.}
\label{analysis_PE_ml-100k1}
\setlength{\tabcolsep}{1.5mm}{
\begin{tabular}{lccccccccc}
\toprule
\multirow{3}{*}{Length} & \multicolumn{9}{c}{\bf \OurMethod{}} \\
\cmidrule(r){2-10}
& \multicolumn{3}{c}{Recall} & \multicolumn{3}{c}{MRR} & \multicolumn{3}{c}{NDCG} \\
\cmidrule(r){2-4}
\cmidrule(r){5-7}
\cmidrule(r){8-10}
& @5 & @10 & @20 & @5 & @10 & @20 & @5 & @10 & @20 \\
\midrule 
$\left[ 20,30 \right)$ & 0.3872 & 0.6011 & 0.7645 & 0.1949 & 0.2232 & 0.2354 & 0.2424 & 0.3113 & 0.3554  \\
$\left[ 30,40 \right)$ & 0.4220 & 0.5871 & 0.7706 & 0.2457 & 0.2670 & 0.2798 & 0.2888 & 0.3414 & 0.3880 \\
$\left[ 40,50 \right]$ & \underline{0.4545} & \underline{0.6590} & \underline{0.7840} & \underline{0.2464} & \underline{0.2713} & \underline{0.2827} & \underline{0.3032} & \underline{0.3628} & \underline{0.4034} \\
$\geq 50$ & 0.2772 & 0.4176 & 0.5857 & 0.1529 & 0.1712 & 0.1827 & 0.1835 & 0.2284 & 0.2708  \\
\bottomrule
\end{tabular}}
\end{table}

\begin{table}[ht]
\centering
\caption{Recall, MRR and NDCG of \OurMethod{} with preference editing on the ``ML-100k'' dataset. The best results are denoted in underline fonts.}
\label{analysis_PE_ml-100k2}
\setlength{\tabcolsep}{1.5mm}{
\begin{tabular}{lccccccccc}
\toprule
\multirow{3}{*}{Length} & \multicolumn{9}{c}{\bf \OurMethod{} (PE)} \\
\cmidrule(r){2-10}
& \multicolumn{3}{c}{Recall} & \multicolumn{3}{c}{MRR} & \multicolumn{3}{c}{NDCG} \\
\cmidrule(r){2-4}
\cmidrule(r){5-7}
\cmidrule(r){8-10}
& @5 & @10 & @20 & @5 & @10 & @20 & @5 & @10 & @20 \\
\midrule 
$\left[ 20,30 \right)$ & 0.4335 & 0.6127 & 0.7861 & 0.2430 & 0.2668 & 0.2788 & \underline{0.2927} & 0.3516 & \underline{0.3954}  \\
$\left[ 30,40 \right)$ & 0.4403 & 0.6238 & 0.7798 & \underline{0.2441} & \underline{0.2694} & \underline{0.2793} & \underline{0.2927} & \underline{0.3528} & 0.3896 \\
$\left[ 40,50 \right]$ & \underline{0.4785} & \underline{0.6785} & \underline{0.8095} & 0.2333 & 0.2544 & 0.2668 & 0.2924 & 0.3462 & 0.3953 \\
$\geq 50$ & 0.3032 & 0.4228 & 0.5875 & 0.1604 & 0.1772 & 0.1887 & 0.1955 & 0.2350 & 0.2767  \\
\bottomrule
\end{tabular}}
\end{table}

\begin{table}[ht]
\centering
\caption{Recall, MRR and NDCG of \OurMethod{} without preference editing on the ``Yelp'' dataset. The best results are denoted in underline fonts.}
\label{analysis_PE_yelp1}
\setlength{\tabcolsep}{1.5mm}{
\begin{tabular}{lccccccccc}
\toprule
\multirow{3}{*}{Length} & \multicolumn{9}{c}{\bf \OurMethod{}} \\
\cmidrule(r){2-10}
& \multicolumn{3}{c}{Recall} & \multicolumn{3}{c}{MRR} & \multicolumn{3}{c}{NDCG} \\
\cmidrule(r){2-4}
\cmidrule(r){5-7}
\cmidrule(r){8-10}
& @5 & @10 & @20 & @5 & @10 & @20 & @5 & @10 & @20 \\
\midrule 
$\left[ 20,30 \right)$ & \underline{0.4730} & \underline{0.6405} & \underline{0.8035} & \underline{0.2842} & \underline{0.3065} & \underline{0.3178} & \underline{0.3310} & \underline{0.3851} & \underline{0.4263}  \\
$\left[ 30,40 \right)$ & 0.4326 & 0.5879 & 0.7606 & 0.2482 & 0.2693 & 0.2814 & 0.2938 & 0.3444 & 0.3882 \\
$\left[ 40,50 \right]$ & 0.4512 & 0.6017 & 0.7694 & 0.2653 & 0.2845 & 0.2871 & 0.3104 & 0.3562 & 0.3971 \\
$\geq 50$ & 0.3756 & 0.5405 & 0.7135 & 0.2171 & 0.2393 & 0.2512 & 0.2563 & 0.3098 & 0.3534  \\
\bottomrule
\end{tabular}}
\end{table}

\begin{table}[ht]
\centering
\caption{Recall, MRR and NDCG of \OurMethod{} with preference editing on the ``Yelp'' dataset. The best results are denoted in underline fonts.}
\label{analysis_PE_yelp2}
\setlength{\tabcolsep}{1.5mm}{
\begin{tabular}{lccccccccc}
\toprule
\multirow{3}{*}{Length} & \multicolumn{9}{c}{\bf \OurMethod{} (PE)} \\
\cmidrule(r){2-10}
& \multicolumn{3}{c}{Recall} & \multicolumn{3}{c}{MRR} & \multicolumn{3}{c}{NDCG} \\
\cmidrule(r){2-4}
\cmidrule(r){5-7}
\cmidrule(r){8-10}
& @5 & @10 & @20 & @5 & @10 & @20 & @5 & @10 & @20 \\
\midrule 
$\left[ 20,30 \right)$ & \underline{0.4840} & \underline{0.6637} & \underline{0.8139} & \underline{0.2950} & \underline{0.3193} & \underline{0.3300} & \underline{0.3418} & \underline{0.4001} & \underline{0.4385}  \\
$\left[ 30,40 \right)$ & 0.4437 & 0.6371 & 0.8035 & 0.2569 & 0.2832 & 0.2947 & 0.3028 & 0.3658 & 0.4078 \\
$\left[ 40,50 \right]$ & 0.4675 & 0.6234 & 0.7792 & 0.2661 & 0.2864 & 0.2948 & 0.3155 & 0.3655 & 0.4045 \\
$\geq 50$ & 0.3946 & 0.5757 & 0.7703 & 0.2309 & 0.2547 & 0.2679 & 0.2714 & 0.3297 & 0.3786  \\
\bottomrule
\end{tabular}}
\end{table}



From these tables, we see that \OurMethod{} (PE) outperforms \OurMethod{} in terms of most metrics over all sequence length groups on all three datasets.
This verifies the effectiveness of preference editing.
On the ``Sports'' dataset, \OurMethod{} (PE) achieves the largest increase when the sequence length is between 30 and 40.
It gains 14.91\%, 9.71\% and 12.08\% in terms of Recall@5, MRR@5 and NDCG@20, respectively.
On the ``ML-100k'' dataset, when the sequence length is between 20 and 30, \OurMethod{} (PE) achieves the largest increase.
It gains 11.95\%, 24.67\% and 20.75\% in terms of Recall@5, MRR@5 and NDCG@5, respectively.
This indicates that preference editing is more effective for sequences which are less than 50 in length.
We also see that the best performances are achieved in different length groups on different datasets.
It achieves the best performances in the groups of [30, 40), [40, 50), [20, 30) on the ``Sports'', ``ML-100k'', ``Yelp'' datasets, respectively.
We believe that this is because of the varied length distributions on different datasets.
From Table~\ref{dataset}, we can see that the average length of ``Sports'' and ``Yelp'' datasets are shorter than that of the ``ML-100k'' dataset.
Hence the best improvements are achieved for longer sequences with a length of no more than 50 on the ``ML-100k'' dataset.
Note that when the sequence length is beyond 50, the performance drops sharply on all three datasets.
This reveals the limitation of \OurMethod{} for very long sequences.

\subsection{Effect of preference coverage}
\label{effect_cover}
To answer RQ3, we remove the coverage mechanism from \OurMethod{} (PE) and keep other settings unchanged.
The results of \OurMethod{} (PE) with/without coverage mechanism are shown in Table~\ref{analysis_coverage}.

\begin{table}[ht]
\centering
\caption{Recall, MRR and NDCG of \OurMethod{} (PE) with and without coverage mechanism on five datasets. Bold face indicates the better results.
}
\label{analysis_coverage}
\setlength{\tabcolsep}{1.0mm}{
\begin{tabular}{l l c c c c c c c c c}
\toprule
\multicolumn{2}{c}{\multirow{2}*{\OurMethod{} (PE)}} & \multicolumn{3}{c}{Recall} & \multicolumn{3}{c}{MRR} & \multicolumn{3}{c}{NDCG} \\
\cmidrule(r){3-5}
\cmidrule(r){6-8}
\cmidrule(r){9-11}
\multicolumn{2}{c}{~} & @5 & @10 & @20 & @5 & @10 & @20 & @5 & @10 & @20 \\
\midrule
\multirow{2}*{Beauty} & -cover & 0.2169 & 0.2913 & 0.4012 & 0.1421 & 0.1520 & 0.1591 & 0.1606 & 0.1846 & 0.2111\\
\cmidrule{2-11}
& +cover & \textbf{0.2286} & \textbf{0.3126} & \textbf{0.4302} & \textbf{0.1444} & \textbf{0.1555} & \textbf{0.1635} & \textbf{0.1653} & \textbf{0.1923} & \textbf{0.2218} \\
\midrule
\multirow{2}*{Sports} & -cover & 0.2263 & 0.3379 & 0.4897 & 0.1309 & 0.1457 & 0.1560 & 0.1545 & 0.1904 & 0.2286 \\
\cmidrule{2-11}
& +cover & \textbf{0.2320} & \textbf{0.3456} & \textbf{0.5000} & \textbf{0.1354} & \textbf{0.1504} & \textbf{0.1616} & \textbf{0.1593} & \textbf{0.1958} & \textbf{0.2347} \\
\midrule
\multirow{2}*{Toys} & -cover & 0.3251 & 0.4191 & 0.5428 & 0.2176 & 0.2301 & 0.2385 & 0.2443 & 0.2746 & 0.3057 \\
\cmidrule{2-11}
& +cover & \textbf{0.3285} & \textbf{0.4218} & \textbf{0.5445} & \textbf{0.2208} & \textbf{0.2332} & \textbf{0.2415} & \textbf{0.2476} & \textbf{0.2776} & \textbf{0.3084} \\
\midrule
\multirow{2}*{ML-100k} & -cover & 0.3351 & 0.4984 & 0.6702 & 0.1916 & 0.2131 & 0.2248 & 0.2270 & 0.2796 & \textbf{0.3227} \\
\cmidrule{2-11}
& +cover & \textbf{0.3601} & \textbf{0.5005} & \textbf{0.6808} & \textbf{0.2031} & \textbf{0.2211} & \textbf{0.2336} & \textbf{0.2416} & \textbf{0.2866} & 0.3221 \\
\midrule
\multirow{2}*{Yelp} & -cover & 0.4579 & 0.6268 & 0.7901 & 0.2729 & 0.2955 & 0.3069 & 0.3187 & 0.3733 & 0.4147 \\
\cmidrule{2-11}
& +cover & \textbf{0.4640} & \textbf{0.6312} & \textbf{0.7972} & \textbf{0.2742} & \textbf{0.2965} & \textbf{0.3080} & \textbf{0.3211} & \textbf{0.3752} & \textbf{0.4172} \\
\bottomrule
\end{tabular}
}
\end{table}

The results of \OurMethod{} (PE) with preference coverage are significantly higher than \OurMethod{} (PE) without preference coverage in terms of most metrics on all datasets.
On the ``Beauty'' dataset, adding preference coverage results in an increase of 7.31\%, 2.76\% and 4.17\% in terms of Recall@10, MRR@20 and NDCG@10, respectively.
And on the ``ML-100k'' dataset, adding preference coverage achieves an increase of 7.46\%, 6.00\% and 6.43\% in terms of Recall@5, MRR@5 and NDCG@5, through NDCG@20 drops a bit.
This is because, without the preference coverage, different preferences might attend to the same items, resulting in redundant preferences and low coverage of the whole sequence.
Besides, the effects of the preference coverage vary on different datasets.
For example, compared with the other datasets, on the ``Toys'' dataset, adding preference coverage results in minor increase of 0.64\%, 1.47\% and 1.35\% in terms of Recall@10, MRR@5 and NDCG@5, respectively.
One possible reason is that the characteristics of the datasets are different, where the number of items interacted by users are smaller and the data is more sparse on ``Toys'' dataset than other datasets.
This may result in little similarity between items.
In this case, the attention distribution in Transformer itself is not very concentrated, so the increase caused by the coverage mechanism is not so large.
What's more, \OurMethod{} (PE) without preference coverage performs worse than some of the baselines on the ``Beauty'' dataset.
This demonstrates the importance and complementarity of preference coverage for preference modeling.
Comparing the results of \OurMethod{} in Table~\ref{analysis_coverage} and Table~\ref{results_1}, we can obtain that removing the preference editing strategy results in worse performances in general than removing the preference coverage mechanism.
It demonstrates that the preference editing strategy plays a leading role in improving recommendation performances.

\subsection{Impact of the hyperparameter $K$}
\label{analysis_K}

\OurMethod{} (PE) concatenates $K$ special tokens to represent user preferences.
Note that $K$ represents latent user preferences for the whole dataset, not for a particular sequence.
To answer RQ4, and study how $K$ affects the recommendation performance of \OurMethod{} (PE), we compare different values of $K$ while keeping other settings unchanged.
We consider $K=1,3,5,7,9,11,15,20$.

\begin{table}[t]
\centering
\caption{Results of \OurMethod{} (PE) on the ``Beauty'' dataset with different values of $K$.}
\setlength{\tabcolsep}{2mm}{
\label{analysis_K_beauty}
\begin{tabular}{@{}r ccccccccc @{}}
\toprule
\multirow{2}{*}{\bf $K$}   & \multicolumn{3}{c}{\bf Recall}      & \multicolumn{3}{c}{\bf MRR}   & \multicolumn{3}{c}{\bf NDCG}         \\ 
\cmidrule(r){2-4} \cmidrule{5-7}  \cmidrule{8-10} & @5     & @10    & @20    & @5      & @10      & @20    & @5     & @10    & @20\\ 
\midrule
1 &  0.2267 & 0.3068  & 0.4185  & 0.1430  & 0.1566  & 0.1642  & \textbf{0.1660} & 0.1918 & 0.2198 \\
3 &  0.2286 & \textbf{0.3126}  & \textbf{0.4302}  & 0.1444 & 0.1555  & 0.1635  & 0.1653 & \textbf{0.1923} &  \textbf{0.2218} \\
5 & 0.2263  & 0.3057 & 0.4217 & 0.1432  & 0.1536  & 0.1615 & 0.1638  & 0.1893 & 0.2184 \\
7 & 0.2250  & 0.3043 & 0.4176  & 0.1422 &  0.1567  & \textbf{0.1644} & 0.1657 & 0.1913  & 0.2197 \\
9 & \textbf{0.2294} & 0.3013 & 0.4170  & \textbf{0.1450}  & \textbf{0.1570} & 0.1637 & 0.1638 & 0.1917 &  0.2215 \\
11 & 0.2237 &  0.3051 & 0.4217  & 0.1441 & 0.1523  & 0.1633  & 0.1642 & 0.1904  & 0.2197 \\
15 & 0.2248  & 0.3021  & 0.4115  &  0.1438  &  0.1550 & 0.1625 & 0.1648 &  0.1895 & 0.2170 \\
20 & 0.2234 & 0.3071 & 0.4173 & 0.1437  & 0.1546 & 0.1621 & 0.1634 & 0.1903  & 0.2180 \\
\bottomrule
\end{tabular}}
\end{table}

\begin{table}[t]
\centering
\caption{Results of \OurMethod{} (PE) on the ``Sports'' dataset with different values of $K$.}
\setlength{\tabcolsep}{2mm}{
\label{analysis_K_sports}
\begin{tabular}{@{}r ccccccccc @{}}
\toprule
\multirow{2}{*}{\bf $K$}   & \multicolumn{3}{c}{\bf Recall}      & \multicolumn{3}{c}{\bf MRR}   & \multicolumn{3}{c}{\bf NDCG}         \\ 
\cmidrule(r){2-4} \cmidrule{5-7}  \cmidrule{8-10} & @5     & @10    & @20    & @5      & @10      & @20    & @5     & @10    & @20\\ 
\midrule
1 &  \textbf{0.2339} & 0.3401  & 0.4812  & \textbf{0.1375}  & \textbf{0.1514}  & 0.1610  & 0.1613 & 0.1954 & 0.2309 \\
3 &  0.2238 & 0.3334  & 0.4840  & 0.1307 & 0.1451  & 0.1554  & 0.1537 & 0.1889 & 0.2268 \\
5 & 0.2283  & 0.3375 & 0.4905 & 0.1324  & 0.1468  & 0.1573 & 0.1560  & 0.1912 & 0.2297 \\
7 & 0.2284  & 0.3397 & 0.4956  & 0.1334 &  0.1480  & 0.1586 & 0.1569 & 0.1926  & 0.2318 \\
9 & 0.2320 & \textbf{0.3456} & \textbf{0.5000}  & 0.1354  & 0.1504 & \textbf{0.1616} & 0.1593 & \textbf{0.1958} &  \textbf{0.2347} \\
11 & 0.2270 &  0.3349 & 0.4865  & 0.1335 & 0.1476  & 0.1579  & 0.1566 & 0.1912  & 0.2293 \\
15 & 0.2285  & 0.3427  & 0.4965  &  0.1334  &  0.1485 & 0.1589 & \textbf{0.1618} &  0.1936 & 0.2322 \\
20 & 0.2175 & 0.3317 & 0.4845 & 0.1259  & 0.1410 & 0.1514 & 0.1485 & 0.1853  & 0.2236 \\
\bottomrule
\end{tabular}}
\end{table}

\begin{table}[t]
\centering
\caption{Results of \OurMethod{} (PE) on the ``Toys'' dataset with different values of $K$.}
\setlength{\tabcolsep}{2mm}{
\label{analysis_K_toys}
\begin{tabular}{@{}r ccccccccc @{}}
\toprule
\multirow{2}{*}{\bf $K$}   & \multicolumn{3}{c}{\bf Recall}      & \multicolumn{3}{c}{\bf MRR}   & \multicolumn{3}{c}{\bf NDCG}         \\ 
\cmidrule(r){2-4} \cmidrule{5-7}  \cmidrule{8-10} & @5     & @10    & @20    & @5      & @10      & @20    & @5     & @10    & @20\\ 
\midrule
1 &  0.3215 & 0.4166  & 0.5356  & 0.2156  & 0.2282  & 0.2363  & 0.2419 & 0.2725 & 0.3025 \\
3 &  0.3230 & 0.4177  & 0.5389  & 0.2171 & 0.2296  & 0.2379  & 0.2434 & 0.2739 &  0.3044 \\
5 & 0.3202  & 0.4162 & 0.5446 & 0.2142  & 0.2269  & 0.2357 & 0.2405  & 0.2715 & 0.3038 \\
7 & 0.3257  & \textbf{0.4243} & \textbf{0.5466}  & 0.2160 &  0.2290  & 0.2374 & 0.2432 & 0.2750  & 0.3037 \\
9 & 0.3123 & 0.4069 & 0.5305  & 0.2044  & 0.2169 & 0.2254 & 0.2312 & 0.2616 &  0.2928 \\
11 & \textbf{0.3285} &  0.4218 & 0.5445  & \textbf{0.2208} & \textbf{0.2332}  & \textbf{0.2415}  & \textbf{0.2476} & \textbf{0.2776}  & \textbf{0.3084} \\
15 & 0.2997  & 0.3982  & 0.5317  &  0.1913  &  0.2043 & 0.2143 & 0.2182 &  0.2499 & 0.2835 \\
20 & 0.3006 & 0.3968 & 0.5208 & 0.1981  & 0.2108 & 0.2193 & 0.2236 & 0.2546  & 0.2857 \\
\bottomrule
\end{tabular}}
\end{table}

\begin{table}
\centering
\caption{Results of \OurMethod{} (PE) on the ``ML-100k'' dataset with different values of $K$.}
\setlength{\tabcolsep}{2mm}{
\label{analysis_K_ml-100k}
\begin{tabular}{@{} r ccccccccc @{}}
\toprule
\multirow{2}{*}{\bf $K$}   & \multicolumn{3}{c}{\bf Recall}      & \multicolumn{3}{c}{\bf MRR}   & \multicolumn{3}{c}{\bf NDCG}\\ 
\cmidrule(r){2-4} \cmidrule{5-7}  \cmidrule{8-10} 
& @5 & @10 & @20 & @5  & @10  & @20  & @5  & @10  & @20\\ 
\midrule
1 &  0.3595 & \textbf{0.5143} & 0.6808  & 0.1975 &  0.2188  & 0.2297 & 0.2375 & \textbf{0.2896} & 0.3297 \\
3 & \textbf{0.3601}  &  0.5005 &  0.6808  & \textbf{0.2031} &  0.2211  & 0.2336 &  \textbf{0.2416} & 0.2866 & 0.3321 \\
5 &  0.3521  & 0.5068 & \textbf{0.6903}  & 0.1991  &  0.2193 &  0.2319 & 0.2367 & 0.2862  & \textbf{0.3325} \\
7 & 0.3383  & 0.4973 &  0.6787  & 0.1948  & 0.2160 & 0.2289 & 0.2303 & 0.2817 & 0.3280 \\
9 & 0.3521 & 0.4942 & 0.6861 & 0.1961  & 0.2154 & 0.2289 & 0.2346 & 0.2809 & 0.3297\\
11 &  0.3351 & 0.4942  & 0.6585  &  0.1940  & 0.2150  & 0.2265 & 0.2287 & 0.2799 &  0.3216 \\
15 &   0.3298 & 0.4835 & 0.6702  & 0.2030 & 0.2211 &  \textbf{0.2341} & 0.2322 & 0.2823 & 0.3294 \\
20 &  0.3552  & 0.5121  & 0.6935  &  0.2010 & \textbf{0.2214}  & 0.2338 & 0.2308 & 0.2790  & 0.3217 \\
\bottomrule
\end{tabular}}
\end{table}

\begin{table}
\centering
\caption{Results of \OurMethod{} (PE) on the ``Yelp'' dataset with different values of $K$.}
\setlength{\tabcolsep}{2mm}{
\label{analysis_K_yelp}
\begin{tabular}{@{} r ccccccccc @{}}
\toprule
\multirow{2}{*}{\bf $K$}   & \multicolumn{3}{c}{\bf Recall}      & \multicolumn{3}{c}{\bf MRR}   & \multicolumn{3}{c}{\bf NDCG}\\ 
\cmidrule(r){2-4} \cmidrule{5-7}  \cmidrule{8-10} 
& @5 & @10 & @20 & @5  & @10  & @20  & @5  & @10  & @20\\ 
\midrule
1 &  \textbf{0.4640} & 0.6312 & 0.7972  & 0.2742 &  \textbf{0.2965}  & \textbf{0.3080} & \textbf{0.3211} & \textbf{0.3752} & \textbf{0.4172} \\
3 & 0.4500  &  0.6284 &  0.8039  & \textbf{0.2787} &  0.2918  & 0.3040 & 0.3154 & 0.3630 & 0.4075 \\
5 &  0.4447  & 0.6173 & 0.7877  & 0.2541  &  0.2771 &  0.2889 & 0.3012 & 0.3570  & 0.4000 \\
7 & 0.4514  & 0.6300 &  0.8012  & 0.2602  & 0.2841 & 0.2960 & 0.3075 & 0.3652 & 0.4086 \\
9 & 0.4385 & 0.6148 & 0.7885 & 0.2490  & 0.2725 & 0.2846 & 0.2958 & 0.3528 & 0.3968\\
11 &  0.4587 & \textbf{0.6361}  & \textbf{0.8055}  &  0.2592  & 0.2830  & 0.2948 & 0.3086 & 0.3660 &  0.4089 \\
15 &  0.4125 & 0.5870 & 0.7634  & 0.2295 & 0.2528 &  0.2651 & 0.2747 & 0.3312 & 0.3758 \\
20 &  0.4354 & 0.6147 & 0.7918  & 0.2488 & 0.2727 &  0.2851 & 0.2949 & 0.3529  & 0.3978 \\
\bottomrule
\end{tabular}}
\end{table}

Table~\ref{analysis_K_beauty}-\ref{analysis_K_yelp} show that the best values of Recall, MRR and NDCG are achieved when $K=3$ on the ``Beauty'' and ``ML-100k'' datasets, $K=9$ on the ``Sports'' dataset, $K=11$ on the ``Toys'' dataset, and $K=1$ on the ``Yelp'' dataset.
This reflects that user preferences on the ``Beauty'', ``ML-100k'' and ``Yelp'' datasets are less extensive than those on the ``Sports'' and ``Toys'' datasets.
Note that this does not conflict with our findings in Section~\ref{rq1} (i.e., the improvements of \OurMethod{} (PE) on ``ML-100k'' dataset are larger than those on other datasets and the advantages of \OurMethod{} (PE) are more obvious on ``ML-100k''), as $K$ represents the number of latent preferences for all sequences in the whole dataset in general, not for a particular sequence.
It is also in line with the number of preferences that users reflect in an interaction sequence in most cases.
The lowest scores are achieved when $K$ is set to a very large value (e.g., $K=15, 20$) on most datasets.
Although the value of $K$ can affect recommendation performance, the influence is limited.
On the ``Beauty'' dataset, the largest gap between the best and the worst performance is 4.54\%, 3.08\% and 2.21\% in terms of Recall@20, MRR@10 and NDCG@20, respectively.


\section{Conclusions and Future Work}
We have proposed a transformer-based model named \OurMethod{} and introduced a novel \ac{SSL} strategy, i.e., \textit{preference editing}. 
We have conducted extensive experiments and analysis on two benchmark datasets to show the effectiveness of \OurMethod{} and preference editing.

Our experimental results demonstrate that \OurMethod{} with preference editing significantly outperforms state-of-the-art methods in terms of Recall, MRR and NDCG.

We have designed \ac{SSL} signals between interaction sequences for the \ac{SR} scenario, that we believe have further uses in the context of recommender systems, including cold-start scenarios and privacy preserving recommendations.

At the same time, \OurMethod{} (PE) also has limitations. Importantly, it is not able to effectively explore user multiple preference from very long sequences.
\OurMethod{} can be advanced and extended in several directions.
First, rich side information can be taken into consideration for user preference extraction and representation, as well as for \ac{SSL}.
Second, variants of \OurMethod{} can be applied to other recommendation tasks by introducing other information, such as conversational recommendations.

\section*{Reproducibility}
To facilitate the reproducibility of the results, we share the datasets, code and parameter files used in this paper at
\my{\url{https://github.com/mamuyang/MrTransformer}}.

\begin{acks}
This research was partially supported by
the National Key R\&D Program of China with grant No. 2020YFB1406704,
the Natural Science Foundation of China (61972234, 61902219, 62072279), 
the Key Scientific and Technological Innovation Program of Shandong Province (2019JZZY010129), 
the Tencent WeChat Rhino-Bird Focused Research Program (JR-WXG-2021411),
the Fundamental Research Funds of Shandong University,
and 
the Hybrid Intelligence Center, a 10-year programme funded by the Dutch Ministry of Education, Culture and Science through the Netherlands Organisation for Scientific Research, \url{https://hybrid-intelligence-centre.nl}.
All content represents the opinion of the authors, which is not necessarily shared or endorsed by their respective employers and/or sponsors.
\end{acks}

\bibliographystyle{ACM-Reference-Format}
\bibliography{main.bbl}


\begin{thebibliography}{71}


\ifx \showCODEN    \undefined \def \showCODEN     #1{\unskip}     \fi
\ifx \showDOI      \undefined \def \showDOI       #1{#1}\fi
\ifx \showISBNx    \undefined \def \showISBNx     #1{\unskip}     \fi
\ifx \showISBNxiii \undefined \def \showISBNxiii  #1{\unskip}     \fi
\ifx \showISSN     \undefined \def \showISSN      #1{\unskip}     \fi
\ifx \showLCCN     \undefined \def \showLCCN      #1{\unskip}     \fi
\ifx \shownote     \undefined \def \shownote      #1{#1}          \fi
\ifx \showarticletitle \undefined \def \showarticletitle #1{#1}   \fi
\ifx \showURL      \undefined \def \showURL       {\relax}        \fi
\providecommand\bibfield[2]{#2}
\providecommand\bibinfo[2]{#2}
\providecommand\natexlab[1]{#1}
\providecommand\showeprint[2][]{arXiv:#2}

\bibitem[\protect\citeauthoryear{Ashkan, Kveton, Berkovsky, and Wen}{Ashkan
  et~al\mbox{.}}{2015}]%
        {diver_ijcai15}
\bibfield{author}{\bibinfo{person}{Azin Ashkan}, \bibinfo{person}{Branislav
  Kveton}, \bibinfo{person}{Shlomo Berkovsky}, {and} \bibinfo{person}{Zheng
  Wen}.} \bibinfo{year}{2015}\natexlab{}.
\newblock \showarticletitle{Optimal greedy diversity for recommendation}. In
  \bibinfo{booktitle}{\emph{The 24th International Joint Conference on
  Artificial Intelligence}}. \bibinfo{pages}{1742--1748}.
\newblock


\bibitem[\protect\citeauthoryear{Chen, Ren, Cai, Sun, and de~Rijke}{Chen
  et~al\mbox{.}}{2020}]%
        {multi-pre-cikm20}
\bibfield{author}{\bibinfo{person}{Wanyu Chen}, \bibinfo{person}{Pengjie Ren},
  \bibinfo{person}{Fei Cai}, \bibinfo{person}{Fei Sun}, {and}
  \bibinfo{person}{Maarten de Rijke}.} \bibinfo{year}{2020}\natexlab{}.
\newblock \showarticletitle{Improving end-to-end sequential recommendations
  with intent-aware diversification}. In \bibinfo{booktitle}{\emph{The 29th
  Conference on Information and Knowledge Management}}.
  \bibinfo{pages}{175--184}.
\newblock


\bibitem[\protect\citeauthoryear{Chen, Xu, Zhang, Tang, Cao, Qin, and Zha}{Chen
  et~al\mbox{.}}{2018}]%
        {chen2018sequential_memNet}
\bibfield{author}{\bibinfo{person}{Xu Chen}, \bibinfo{person}{Hongteng Xu},
  \bibinfo{person}{Yongfeng Zhang}, \bibinfo{person}{Jiaxi Tang},
  \bibinfo{person}{Yixin Cao}, \bibinfo{person}{Zheng Qin}, {and}
  \bibinfo{person}{Hongyuan Zha}.} \bibinfo{year}{2018}\natexlab{}.
\newblock \showarticletitle{Sequential recommendation with user memory
  networks}. In \bibinfo{booktitle}{\emph{The 11th Conferences on Web-inspired
  research involving Search and Data Mining}}. \bibinfo{pages}{108--116}.
\newblock


\bibitem[\protect\citeauthoryear{Cho, Park, and Yoo}{Cho et~al\mbox{.}}{2020}]%
        {cho2020meantime}
\bibfield{author}{\bibinfo{person}{Sung~Min Cho}, \bibinfo{person}{Eunhyeok
  Park}, {and} \bibinfo{person}{Sungjoo Yoo}.} \bibinfo{year}{2020}\natexlab{}.
\newblock \showarticletitle{MEANTIME: Mixture of attention mechanisms with
  multi-temporal embeddings for sequential recommendation}. In
  \bibinfo{booktitle}{\emph{The 14th ACM Conference on Recommender Systems}}.
  \bibinfo{pages}{515--520}.
\newblock


\bibitem[\protect\citeauthoryear{Devlin, Chang, Lee, and Toutanova}{Devlin
  et~al\mbox{.}}{2019}]%
        {nlp_Devlin2019}
\bibfield{author}{\bibinfo{person}{Jacob Devlin}, \bibinfo{person}{Ming-Wei
  Chang}, \bibinfo{person}{Kenton Lee}, {and} \bibinfo{person}{Kristina
  Toutanova}.} \bibinfo{year}{2019}\natexlab{}.
\newblock \showarticletitle{BERT: Pre-training of deep bidirectional
  Transformers for language understanding}. In \bibinfo{booktitle}{\emph{The
  North American Chapter of the Association for Computational Linguistics}}.
  \bibinfo{pages}{4171--4186}.
\newblock


\bibitem[\protect\citeauthoryear{Donkers, Loepp, and Ziegler}{Donkers
  et~al\mbox{.}}{2017}]%
        {donkers2017sequential}
\bibfield{author}{\bibinfo{person}{Tim Donkers}, \bibinfo{person}{Benedikt
  Loepp}, {and} \bibinfo{person}{J{\"u}rgen Ziegler}.}
  \bibinfo{year}{2017}\natexlab{}.
\newblock \showarticletitle{Sequential user-based recurrent neural network
  recommendations}. In \bibinfo{booktitle}{\emph{The 11th ACM Conference on
  Recommender Systems}}. \bibinfo{pages}{152--160}.
\newblock


\bibitem[\protect\citeauthoryear{Felbo, Mislove, S{\o}gaard, Rahwan, and
  Lehmann}{Felbo et~al\mbox{.}}{2017}]%
        {nlp_Felbo2017}
\bibfield{author}{\bibinfo{person}{Bjarke Felbo}, \bibinfo{person}{Alan
  Mislove}, \bibinfo{person}{Anders S{\o}gaard}, \bibinfo{person}{Iyad Rahwan},
  {and} \bibinfo{person}{Sune Lehmann}.} \bibinfo{year}{2017}\natexlab{}.
\newblock \showarticletitle{Using millions of emoji occurrences to learn
  any-domain representations for detecting sentiment, emotion and sarcasm}. In
  \bibinfo{booktitle}{\emph{The Conference on Empirical Methods in Natural
  Language Processing}}. \bibinfo{pages}{1615--1625}.
\newblock


\bibitem[\protect\citeauthoryear{Guo, Zhang, Fang, Jin, and Pan}{Guo
  et~al\mbox{.}}{2020}]%
        {guo2020session}
\bibfield{author}{\bibinfo{person}{Cheng Guo}, \bibinfo{person}{Mengfei Zhang},
  \bibinfo{person}{Jinyun Fang}, \bibinfo{person}{Jiaqi Jin}, {and}
  \bibinfo{person}{Mao Pan}.} \bibinfo{year}{2020}\natexlab{}.
\newblock \showarticletitle{Session-based recommendation with hierarchical
  leaping networks}. In \bibinfo{booktitle}{\emph{The 43rd International ACM
  Conference on Research and Development in Information Retrieval}}.
  \bibinfo{pages}{1705--1708}.
\newblock


\bibitem[\protect\citeauthoryear{He, Fan, Wu, Xie, and Girshick}{He
  et~al\mbox{.}}{2020}]%
        {MoCo2020}
\bibfield{author}{\bibinfo{person}{Kaiming He}, \bibinfo{person}{Haoqi Fan},
  \bibinfo{person}{Yuxin Wu}, \bibinfo{person}{Saining Xie}, {and}
  \bibinfo{person}{Ross Girshick}.} \bibinfo{year}{2020}\natexlab{}.
\newblock \showarticletitle{Momentum contrast for unsupervised visual
  representation learning}. In \bibinfo{booktitle}{\emph{Conference on Computer
  Vision and Pattern Recognition}}. \bibinfo{pages}{9729--9738}.
\newblock


\bibitem[\protect\citeauthoryear{He, Liao, Zhang, Nie, Hu, and Chua}{He
  et~al\mbox{.}}{2017}]%
        {NCF}
\bibfield{author}{\bibinfo{person}{Xiangnan He}, \bibinfo{person}{Lizi Liao},
  \bibinfo{person}{Hanwang Zhang}, \bibinfo{person}{Liqiang Nie},
  \bibinfo{person}{Xia Hu}, {and} \bibinfo{person}{Tat-Seng Chua}.}
  \bibinfo{year}{2017}\natexlab{}.
\newblock \showarticletitle{Neural collaborative filtering}. In
  \bibinfo{booktitle}{\emph{The 26th Web Conference}}.
  \bibinfo{pages}{173--182}.
\newblock


\bibitem[\protect\citeauthoryear{Hidasi, Karatzoglou, Baltrunas, and
  Tikk}{Hidasi et~al\mbox{.}}{2016a}]%
        {GRU4rec}
\bibfield{author}{\bibinfo{person}{Bal{\'{a}}zs Hidasi},
  \bibinfo{person}{Alexandros Karatzoglou}, \bibinfo{person}{Linas Baltrunas},
  {and} \bibinfo{person}{Domonkos Tikk}.} \bibinfo{year}{2016}\natexlab{a}.
\newblock \showarticletitle{Session-based recommendations with recurrent neural
  networks}. In \bibinfo{booktitle}{\emph{The 4th International Conference on
  Learning Representations}}.
\newblock


\bibitem[\protect\citeauthoryear{Hidasi, Quadrana, Karatzoglou, and
  Tikk}{Hidasi et~al\mbox{.}}{2016b}]%
        {hidasi2016parallel}
\bibfield{author}{\bibinfo{person}{Bal{\'a}zs Hidasi}, \bibinfo{person}{Massimo
  Quadrana}, \bibinfo{person}{Alexandros Karatzoglou}, {and}
  \bibinfo{person}{Domonkos Tikk}.} \bibinfo{year}{2016}\natexlab{b}.
\newblock \showarticletitle{Parallel recurrent neural network architectures for
  feature-rich session-based recommendations}. In \bibinfo{booktitle}{\emph{The
  10th ACM Conference on Recommender Systems}}. \bibinfo{pages}{241--248}.
\newblock


\bibitem[\protect\citeauthoryear{Huang, Ren, Zhao, He, Wen, and Dong}{Huang
  et~al\mbox{.}}{2019}]%
        {huang2019taxonomy}
\bibfield{author}{\bibinfo{person}{Jin Huang}, \bibinfo{person}{Zhaochun Ren},
  \bibinfo{person}{Wayne~Xin Zhao}, \bibinfo{person}{Gaole He},
  \bibinfo{person}{Ji-Rong Wen}, {and} \bibinfo{person}{Daxiang Dong}.}
  \bibinfo{year}{2019}\natexlab{}.
\newblock \showarticletitle{Taxonomy-aware multi-hop reasoning networks for
  sequential recommendation}. In \bibinfo{booktitle}{\emph{The 12th Conferences
  on Web-inspired research involving Search and Data Mining}}.
  \bibinfo{pages}{573--581}.
\newblock


\bibitem[\protect\citeauthoryear{Huang, Zhao, Dou, Wen, and Chang}{Huang
  et~al\mbox{.}}{2018}]%
        {KSR}
\bibfield{author}{\bibinfo{person}{Jin Huang}, \bibinfo{person}{Wayne~Xin
  Zhao}, \bibinfo{person}{Hongjian Dou}, \bibinfo{person}{Ji-Rong Wen}, {and}
  \bibinfo{person}{Edward~Y Chang}.} \bibinfo{year}{2018}\natexlab{}.
\newblock \showarticletitle{Improving sequential recommendation with
  knowledge-enhanced memory networks}. In \bibinfo{booktitle}{\emph{The 41st
  International ACM Conference on Research and Development in Information
  Retrieval}}. \bibinfo{pages}{505--514}.
\newblock


\bibitem[\protect\citeauthoryear{Kang and McAuley}{Kang and McAuley}{2018}]%
        {SASRec}
\bibfield{author}{\bibinfo{person}{Wangcheng Kang} {and}
  \bibinfo{person}{Julian McAuley}.} \bibinfo{year}{2018}\natexlab{}.
\newblock \showarticletitle{Self-attentive sequential recommendation}. In
  \bibinfo{booktitle}{\emph{The 18th International Conference on Data Mining}}.
  \bibinfo{pages}{197--206}.
\newblock


\bibitem[\protect\citeauthoryear{Lan, Chen, Goodman, Gimpel, Sharma, and
  Soricut}{Lan et~al\mbox{.}}{2020}]%
        {nlp_Lan2020}
\bibfield{author}{\bibinfo{person}{Zhenzhong Lan}, \bibinfo{person}{Mingda
  Chen}, \bibinfo{person}{Sebastian Goodman}, \bibinfo{person}{Kevin Gimpel},
  \bibinfo{person}{Piyush Sharma}, {and} \bibinfo{person}{Radu Soricut}.}
  \bibinfo{year}{2020}\natexlab{}.
\newblock \showarticletitle{ALBERT: A lite BERT for self-supervised learning of
  language representations}. In \bibinfo{booktitle}{\emph{The 8th International
  Conference on Learning Representations}}.
\newblock


\bibitem[\protect\citeauthoryear{Lewis, Liu, Goyal, Ghazvininejad, Mohamed,
  Levy, Stoyanov, and Zettlemoyer}{Lewis et~al\mbox{.}}{2020}]%
        {nlp_Lewis2020}
\bibfield{author}{\bibinfo{person}{Mike Lewis}, \bibinfo{person}{Yinhan Liu},
  \bibinfo{person}{Naman Goyal}, \bibinfo{person}{Marjan Ghazvininejad},
  \bibinfo{person}{Abdelrahman Mohamed}, \bibinfo{person}{Omer Levy},
  \bibinfo{person}{Ves Stoyanov}, {and} \bibinfo{person}{Luke Zettlemoyer}.}
  \bibinfo{year}{2020}\natexlab{}.
\newblock \showarticletitle{BART: Denoising sequence-to-sequence pre-training
  for natural language generation, translation, and comprehension}. In
  \bibinfo{booktitle}{\emph{The 58th Association for Computational
  Linguistics}}. \bibinfo{pages}{7871--7880}.
\newblock


\bibitem[\protect\citeauthoryear{Li, Niu, Luo, Chen, and Quan}{Li
  et~al\mbox{.}}{2019}]%
        {review_driven_sr2019}
\bibfield{author}{\bibinfo{person}{Chenliang Li}, \bibinfo{person}{Xichuan
  Niu}, \bibinfo{person}{Xiangyang Luo}, \bibinfo{person}{Zhenzhong Chen},
  {and} \bibinfo{person}{Cong Quan}.} \bibinfo{year}{2019}\natexlab{}.
\newblock \showarticletitle{A review-driven neural model for sequential
  recommendation}. In \bibinfo{booktitle}{\emph{The 28th International Joint
  Conference on Artificial Intelligence}}. \bibinfo{pages}{2866--2872}.
\newblock


\bibitem[\protect\citeauthoryear{Li, Ren, Chen, Ren, Lian, and Ma}{Li
  et~al\mbox{.}}{2017}]%
        {NARM}
\bibfield{author}{\bibinfo{person}{Jing Li}, \bibinfo{person}{Pengjie Ren},
  \bibinfo{person}{Zhumin Chen}, \bibinfo{person}{Zhaochun Ren},
  \bibinfo{person}{Tao Lian}, {and} \bibinfo{person}{Jun Ma}.}
  \bibinfo{year}{2017}\natexlab{}.
\newblock \showarticletitle{Neural attentive session-based recommendation}. In
  \bibinfo{booktitle}{\emph{The 26th Conference on Information and Knowledge
  Management}}. \bibinfo{pages}{1419--1428}.
\newblock


\bibitem[\protect\citeauthoryear{Li, Wang, and McAuley}{Li
  et~al\mbox{.}}{2020}]%
        {li2020time}
\bibfield{author}{\bibinfo{person}{Jiacheng Li}, \bibinfo{person}{Yujie Wang},
  {and} \bibinfo{person}{Julian McAuley}.} \bibinfo{year}{2020}\natexlab{}.
\newblock \showarticletitle{Time interval aware self-attention for sequential
  recommendation}. In \bibinfo{booktitle}{\emph{The 13th Conferences on
  Web-inspired research involving Search and Data Mining}}.
  \bibinfo{pages}{322--330}.
\newblock


\bibitem[\protect\citeauthoryear{Lin, Pan, and Ming}{Lin et~al\mbox{.}}{2020}]%
        {lin2020fissa}
\bibfield{author}{\bibinfo{person}{Jing Lin}, \bibinfo{person}{Weike Pan},
  {and} \bibinfo{person}{Zhong Ming}.} \bibinfo{year}{2020}\natexlab{}.
\newblock \showarticletitle{FISSA: Fusing item similarity models with
  self-attention networks for sequential recommendation}. In
  \bibinfo{booktitle}{\emph{The 14th ACM Conference on Recommender Systems}}.
  \bibinfo{pages}{130--139}.
\newblock


\bibitem[\protect\citeauthoryear{Linden, Smith, and York}{Linden
  et~al\mbox{.}}{2003}]%
        {itemknn}
\bibfield{author}{\bibinfo{person}{Greg Linden}, \bibinfo{person}{Brent Smith},
  {and} \bibinfo{person}{Jeremy York}.} \bibinfo{year}{2003}\natexlab{}.
\newblock \showarticletitle{Amazon.com recommendations: item-to-item
  collaborative filtering}.
\newblock \bibinfo{journal}{\emph{IEEE Internet Computing}}
  \bibinfo{volume}{1} (\bibinfo{year}{2003}), \bibinfo{pages}{76--80}.
\newblock


\bibitem[\protect\citeauthoryear{Liu, Li, Cai, Dong, Shang, and Zhu}{Liu
  et~al\mbox{.}}{2021a}]%
        {Noninvasive}
\bibfield{author}{\bibinfo{person}{Chang Liu}, \bibinfo{person}{Xiaoguang Li},
  \bibinfo{person}{Guohao Cai}, \bibinfo{person}{Zhenhua Dong},
  \bibinfo{person}{Lifeng Shang}, {and} \bibinfo{person}{Hong Zhu}.}
  \bibinfo{year}{2021}\natexlab{a}.
\newblock \showarticletitle{Noninvasive self-attention for side information
  fusion in sequential recommendation}. In \bibinfo{booktitle}{\emph{The 35th
  Conference on Artificial Intelligence}}.
\newblock


\bibitem[\protect\citeauthoryear{Liu, Zeng, Mokhosi, and Zhang}{Liu
  et~al\mbox{.}}{2018}]%
        {STAMP}
\bibfield{author}{\bibinfo{person}{Qiao Liu}, \bibinfo{person}{Yifu Zeng},
  \bibinfo{person}{Refuoe Mokhosi}, {and} \bibinfo{person}{Haibin Zhang}.}
  \bibinfo{year}{2018}\natexlab{}.
\newblock \showarticletitle{STAMP: Short-term attention/memory priority model
  for session-based recommendation}. In \bibinfo{booktitle}{\emph{The 24th ACM
  International Conference on Knowledge discovery and data mining}}.
  \bibinfo{pages}{1831--1839}.
\newblock


\bibitem[\protect\citeauthoryear{Liu, Ma, Ouyang, and Xiong}{Liu
  et~al\mbox{.}}{2021b}]%
        {liu2021contrastive}
\bibfield{author}{\bibinfo{person}{Zhuang Liu}, \bibinfo{person}{Yunpu Ma},
  \bibinfo{person}{Yuanxin Ouyang}, {and} \bibinfo{person}{Zhang Xiong}.}
  \bibinfo{year}{2021}\natexlab{b}.
\newblock \showarticletitle{Contrastive learning for recommender system}. In
  \bibinfo{booktitle}{\emph{arXiv preprint arXiv:2101.01317}}.
\newblock


\bibitem[\protect\citeauthoryear{Luo, Zhao, Liu, Zhuang, Wang, Xu, Fang, and
  Sheng}{Luo et~al\mbox{.}}{2020}]%
        {luocollaborative}
\bibfield{author}{\bibinfo{person}{Anjing Luo}, \bibinfo{person}{Pengpeng
  Zhao}, \bibinfo{person}{Yanchi Liu}, \bibinfo{person}{Fuzhen Zhuang},
  \bibinfo{person}{Deqing Wang}, \bibinfo{person}{Jiajie Xu},
  \bibinfo{person}{Junhua Fang}, {and} \bibinfo{person}{Victor~S Sheng}.}
  \bibinfo{year}{2020}\natexlab{}.
\newblock \showarticletitle{Collaborative Self-Attention Network for
  Session-based Recommendation}. In \bibinfo{booktitle}{\emph{The 29th
  International Joint Conference on Artificial Intelligence}}.
  \bibinfo{pages}{2591--2597}.
\newblock


\bibitem[\protect\citeauthoryear{Ma, Zhou, Yang, Cui, Wang, and Zhu}{Ma
  et~al\mbox{.}}{2020}]%
        {ssl_cl_s2s2020}
\bibfield{author}{\bibinfo{person}{Jianxin Ma}, \bibinfo{person}{Chang Zhou},
  \bibinfo{person}{Hongxia Yang}, \bibinfo{person}{Peng Cui},
  \bibinfo{person}{Xin Wang}, {and} \bibinfo{person}{Wenwu Zhu}.}
  \bibinfo{year}{2020}\natexlab{}.
\newblock \showarticletitle{Disentangled self-Supervision in sequential
  recommenders}. In \bibinfo{booktitle}{\emph{The 26th ACM International
  Conference on Knowledge discovery and data mining}}.
  \bibinfo{pages}{483--491}.
\newblock


\bibitem[\protect\citeauthoryear{Ma, Ren, Lin, Chen, Ma, and de~Rijke}{Ma
  et~al\mbox{.}}{2019}]%
        {ma2019pi}
\bibfield{author}{\bibinfo{person}{Muyang Ma}, \bibinfo{person}{Pengjie Ren},
  \bibinfo{person}{Yujie Lin}, \bibinfo{person}{Zhumin Chen},
  \bibinfo{person}{Jun Ma}, {and} \bibinfo{person}{Maarten de Rijke}.}
  \bibinfo{year}{2019}\natexlab{}.
\newblock \showarticletitle{$\pi$-Net: A Parallel information-sharing network
  for shared-account cross-domain sequential recommendations}. In
  \bibinfo{booktitle}{\emph{The 42nd International ACM Conference on Research
  and Development in Information Retrieval}}. \bibinfo{pages}{685--694}.
\newblock


\bibitem[\protect\citeauthoryear{McAuley, Targett, Shi, and Van
  Den~Hengel}{McAuley et~al\mbox{.}}{2015}]%
        {mcauley2015image}
\bibfield{author}{\bibinfo{person}{Julian McAuley},
  \bibinfo{person}{Christopher Targett}, \bibinfo{person}{Qinfeng Shi}, {and}
  \bibinfo{person}{Anton Van Den~Hengel}.} \bibinfo{year}{2015}\natexlab{}.
\newblock \showarticletitle{Image-based recommendations on styles and
  substitutes}. In \bibinfo{booktitle}{\emph{The 38th International ACM
  Conference on Research and Development in Information Retrieval}}.
  \bibinfo{pages}{43--52}.
\newblock


\bibitem[\protect\citeauthoryear{Meng, Yang, and Xiao}{Meng
  et~al\mbox{.}}{2020}]%
        {meng2020incorporating}
\bibfield{author}{\bibinfo{person}{Wenjing Meng}, \bibinfo{person}{Deqing
  Yang}, {and} \bibinfo{person}{Yanghua Xiao}.}
  \bibinfo{year}{2020}\natexlab{}.
\newblock \showarticletitle{Incorporating user micro-behaviors and item
  knowledge into multi-task learning for session-based recommendation}. In
  \bibinfo{booktitle}{\emph{The 43rd International ACM Conference on Research
  and Development in Information Retrieval}}. \bibinfo{pages}{1091--1100}.
\newblock


\bibitem[\protect\citeauthoryear{Mi, Lin, and Faltings}{Mi
  et~al\mbox{.}}{2020}]%
        {mi2020ader}
\bibfield{author}{\bibinfo{person}{Fei Mi}, \bibinfo{person}{Xiaoyu Lin}, {and}
  \bibinfo{person}{Boi Faltings}.} \bibinfo{year}{2020}\natexlab{}.
\newblock \showarticletitle{Ader: Adaptively distilled exemplar replay towards
  continual learning for session-based recommendation}. In
  \bibinfo{booktitle}{\emph{The 14th ACM Conference on Recommender Systems}}.
  \bibinfo{pages}{408--413}.
\newblock


\bibitem[\protect\citeauthoryear{Pan, Cai, Ling, and de~Rijke}{Pan
  et~al\mbox{.}}{2020a}]%
        {pan2020intent}
\bibfield{author}{\bibinfo{person}{Zhiqiang Pan}, \bibinfo{person}{Fei Cai},
  \bibinfo{person}{Yanxiang Ling}, {and} \bibinfo{person}{Maarten de Rijke}.}
  \bibinfo{year}{2020}\natexlab{a}.
\newblock \showarticletitle{An intent-guided collaborative machine for
  session-based recommendation}. In \bibinfo{booktitle}{\emph{The 43rd
  International ACM Conference on Research and Development in Information
  Retrieval}}. \bibinfo{pages}{1833--1836}.
\newblock


\bibitem[\protect\citeauthoryear{Pan, Cai, Ling, and de~Rijke}{Pan
  et~al\mbox{.}}{2020b}]%
        {pan2020rethinking}
\bibfield{author}{\bibinfo{person}{Zhiqiang Pan}, \bibinfo{person}{Fei Cai},
  \bibinfo{person}{Yanxiang Ling}, {and} \bibinfo{person}{Maarten de Rijke}.}
  \bibinfo{year}{2020}\natexlab{b}.
\newblock \showarticletitle{Rethinking item importance in session-based
  recommendation}. In \bibinfo{booktitle}{\emph{The 43rd International ACM
  Conference on Research and Development in Information Retrieval}}.
  \bibinfo{pages}{1837--1840}.
\newblock


\bibitem[\protect\citeauthoryear{Qiu, Yin, Huang, and Chen}{Qiu
  et~al\mbox{.}}{2020}]%
        {qiu2020gag}
\bibfield{author}{\bibinfo{person}{Ruihong Qiu}, \bibinfo{person}{Hongzhi Yin},
  \bibinfo{person}{Zi Huang}, {and} \bibinfo{person}{Tong Chen}.}
  \bibinfo{year}{2020}\natexlab{}.
\newblock \showarticletitle{GAG: Global attributed graph neural network for
  streaming session-based recommendation}. In \bibinfo{booktitle}{\emph{The
  43rd International ACM Conference on Research and Development in Information
  Retrieval}}. \bibinfo{pages}{669--678}.
\newblock


\bibitem[\protect\citeauthoryear{Quadrana, Karatzoglou, Hidasi, and
  Cremonesi}{Quadrana et~al\mbox{.}}{2017}]%
        {HRNN}
\bibfield{author}{\bibinfo{person}{Massimo Quadrana},
  \bibinfo{person}{Alexandros Karatzoglou}, \bibinfo{person}{Balzs Hidasi},
  {and} \bibinfo{person}{Paolo Cremonesi}.} \bibinfo{year}{2017}\natexlab{}.
\newblock \showarticletitle{Personalizing session-based recommendations with
  hierarchical recurrent neural networks}. In \bibinfo{booktitle}{\emph{The
  11th ACM Conference on Recommender Systems}}. \bibinfo{pages}{130--137}.
\newblock


\bibitem[\protect\citeauthoryear{Ren, Chen, Li, Ren, Ma, and de~Rijke}{Ren
  et~al\mbox{.}}{2019}]%
        {repeatnet}
\bibfield{author}{\bibinfo{person}{Pengjie Ren}, \bibinfo{person}{Zhumin Chen},
  \bibinfo{person}{Jing Li}, \bibinfo{person}{Zhaochun Ren},
  \bibinfo{person}{Jun Ma}, {and} \bibinfo{person}{Maarten de Rijke}.}
  \bibinfo{year}{2019}\natexlab{}.
\newblock \showarticletitle{RepeatNet: A repeat aware neural recommendation
  machine for session-based recommendation}. In \bibinfo{booktitle}{\emph{The
  33rd Conference on Artificial Intelligence}}. \bibinfo{pages}{4806--4813}.
\newblock


\bibitem[\protect\citeauthoryear{Ren, Ren, Sun, He, Yin, and de~Rijke}{Ren
  et~al\mbox{.}}{2020b}]%
        {ren2020nlp4rec}
\bibfield{author}{\bibinfo{person}{Pengjie Ren}, \bibinfo{person}{Zhaochun
  Ren}, \bibinfo{person}{Fei Sun}, \bibinfo{person}{Xiangnan He},
  \bibinfo{person}{Dawei Yin}, {and} \bibinfo{person}{Maarten de Rijke}.}
  \bibinfo{year}{2020}\natexlab{b}.
\newblock \showarticletitle{NLP4REC: The WSDM 2020 workshop on natural language
  processing for recommendations}. In \bibinfo{booktitle}{\emph{The 13th
  Conferences on Web-inspired research involving Search and Data Mining}}.
  \bibinfo{pages}{907--908}.
\newblock


\bibitem[\protect\citeauthoryear{Ren, Liu, Li, Zhao, Wang, Ding, and Wen}{Ren
  et~al\mbox{.}}{2020a}]%
        {masr2020}
\bibfield{author}{\bibinfo{person}{Ruiyang Ren}, \bibinfo{person}{Zhaoyang
  Liu}, \bibinfo{person}{Yaliang Li}, \bibinfo{person}{Wayne~Xin Zhao},
  \bibinfo{person}{Hui Wang}, \bibinfo{person}{Bolin Ding}, {and}
  \bibinfo{person}{Ji{-}Rong Wen}.} \bibinfo{year}{2020}\natexlab{a}.
\newblock \showarticletitle{Sequential recommendation with self-attentive
  multi-adversarial network}. In \bibinfo{booktitle}{\emph{The 43rd
  International ACM Conference on Research and Development in Information
  Retrieval}}. \bibinfo{pages}{89--98}.
\newblock


\bibitem[\protect\citeauthoryear{Song, Xiao, Wang, Charlin, Zhang, and
  Tang}{Song et~al\mbox{.}}{2019}]%
        {song2019session}
\bibfield{author}{\bibinfo{person}{Weiping Song}, \bibinfo{person}{Zhiping
  Xiao}, \bibinfo{person}{Yifan Wang}, \bibinfo{person}{Laurent Charlin},
  \bibinfo{person}{Ming Zhang}, {and} \bibinfo{person}{Jian Tang}.}
  \bibinfo{year}{2019}\natexlab{}.
\newblock \showarticletitle{Session-based social recommendation via dynamic
  graph attention networks}. In \bibinfo{booktitle}{\emph{The 12th Conferences
  on Web-inspired research involving Search and Data Mining}}.
  \bibinfo{pages}{555--563}.
\newblock


\bibitem[\protect\citeauthoryear{Sun, Liu, Wu, Pei, Lin, Ou, and Jiang}{Sun
  et~al\mbox{.}}{2019a}]%
        {bert4rec}
\bibfield{author}{\bibinfo{person}{Fei Sun}, \bibinfo{person}{Jun Liu},
  \bibinfo{person}{Jian Wu}, \bibinfo{person}{Changhua Pei},
  \bibinfo{person}{Xiao Lin}, \bibinfo{person}{Wenwu Ou}, {and}
  \bibinfo{person}{Peng Jiang}.} \bibinfo{year}{2019}\natexlab{a}.
\newblock \showarticletitle{BERT4Rec: Sequential recommendation with
  bidirectional encoder representations from Transformer}. In
  \bibinfo{booktitle}{\emph{The 28th Conference on Information and Knowledge
  Management}}. \bibinfo{pages}{1441--1450}.
\newblock


\bibitem[\protect\citeauthoryear{Sun, Wu, and Wang}{Sun et~al\mbox{.}}{2018}]%
        {sun2018attentive}
\bibfield{author}{\bibinfo{person}{Peijie Sun}, \bibinfo{person}{Le Wu}, {and}
  \bibinfo{person}{Meng Wang}.} \bibinfo{year}{2018}\natexlab{}.
\newblock \showarticletitle{Attentive recurrent social recommendation}. In
  \bibinfo{booktitle}{\emph{The 41st International ACM Conference on Research
  and Development in Information Retrieval}}. \bibinfo{pages}{185--194}.
\newblock


\bibitem[\protect\citeauthoryear{Sun, Wu, Zhang, Fu, Hong, and Wang}{Sun
  et~al\mbox{.}}{2020a}]%
        {sun2020dual}
\bibfield{author}{\bibinfo{person}{Peijie Sun}, \bibinfo{person}{Le Wu},
  \bibinfo{person}{Kun Zhang}, \bibinfo{person}{Yanjie Fu},
  \bibinfo{person}{Richang Hong}, {and} \bibinfo{person}{Meng Wang}.}
  \bibinfo{year}{2020}\natexlab{a}.
\newblock \showarticletitle{Dual learning for explainable recommendation:
  Towards unifying user preference prediction and review generation}. In
  \bibinfo{booktitle}{\emph{The 29th Web Conference}}.
  \bibinfo{pages}{837--847}.
\newblock


\bibitem[\protect\citeauthoryear{Sun, Tzeng, Darrell, and A.~Efros}{Sun
  et~al\mbox{.}}{2019b}]%
        {cv_sun2019}
\bibfield{author}{\bibinfo{person}{Yu Sun}, \bibinfo{person}{Eric Tzeng},
  \bibinfo{person}{Trevor Darrell}, {and} \bibinfo{person}{Alexei A.~Efros}.}
  \bibinfo{year}{2019}\natexlab{b}.
\newblock \showarticletitle{Unsupervised domain adaptation through
  self-supervision}. In \bibinfo{booktitle}{\emph{arXiv preprint
  arXiv:1909.11825}}.
\newblock


\bibitem[\protect\citeauthoryear{Sun, Yuan, Yang, Wei, Zhao, and Liu}{Sun
  et~al\mbox{.}}{2020b}]%
        {sun2020generic}
\bibfield{author}{\bibinfo{person}{Yang Sun}, \bibinfo{person}{Fajie Yuan},
  \bibinfo{person}{Min Yang}, \bibinfo{person}{Guoao Wei},
  \bibinfo{person}{Zhou Zhao}, {and} \bibinfo{person}{Duo Liu}.}
  \bibinfo{year}{2020}\natexlab{b}.
\newblock \showarticletitle{A generic network compression framework for
  sequential recommender systems}. In \bibinfo{booktitle}{\emph{The 43rd
  International ACM Conference on Research and Development in Information
  Retrieval}}. \bibinfo{pages}{1299--1308}.
\newblock


\bibitem[\protect\citeauthoryear{Tang and Wang}{Tang and Wang}{2018}]%
        {Caser2018}
\bibfield{author}{\bibinfo{person}{Jiaxi Tang} {and} \bibinfo{person}{Ke
  Wang}.} \bibinfo{year}{2018}\natexlab{}.
\newblock \showarticletitle{Personalized top-n sequential recommendation via
  convolutional sequence embedding}. In \bibinfo{booktitle}{\emph{The 11th
  Conferences on Web-inspired research involving Search and Data Mining}}.
  \bibinfo{pages}{565--573}.
\newblock


\bibitem[\protect\citeauthoryear{van~den Oord, Li, and Vinyals}{van~den Oord
  et~al\mbox{.}}{2018}]%
        {cv_vandenOord2018}
\bibfield{author}{\bibinfo{person}{Aaron van~den Oord}, \bibinfo{person}{Yazhe
  Li}, {and} \bibinfo{person}{Oriol Vinyals}.} \bibinfo{year}{2018}\natexlab{}.
\newblock \showarticletitle{Representation learning with contrastive predictive
  coding}. In \bibinfo{booktitle}{\emph{arXiv preprint arXiv:1807.03748}}.
\newblock


\bibitem[\protect\citeauthoryear{Wang, Ding, Hong, Liu, and Caverlee}{Wang
  et~al\mbox{.}}{2020a}]%
        {wang2020next}
\bibfield{author}{\bibinfo{person}{Jianling Wang}, \bibinfo{person}{Kaize
  Ding}, \bibinfo{person}{Liangjie Hong}, \bibinfo{person}{Huan Liu}, {and}
  \bibinfo{person}{James Caverlee}.} \bibinfo{year}{2020}\natexlab{a}.
\newblock \showarticletitle{Next-item recommendation with sequential
  hypergraphs}. In \bibinfo{booktitle}{\emph{The 43rd International ACM
  Conference on Research and Development in Information Retrieval}}.
  \bibinfo{pages}{1101--1110}.
\newblock


\bibitem[\protect\citeauthoryear{Wang, Fu, He, Hao, and Wu}{Wang
  et~al\mbox{.}}{2020c}]%
        {wang2020survey}
\bibfield{author}{\bibinfo{person}{Meng Wang}, \bibinfo{person}{Weijie Fu},
  \bibinfo{person}{Xiangnan He}, \bibinfo{person}{Shijie Hao}, {and}
  \bibinfo{person}{Xindong Wu}.} \bibinfo{year}{2020}\natexlab{c}.
\newblock \showarticletitle{A Survey on Large-scale Machine Learning}.
\newblock  (\bibinfo{year}{2020}).
\newblock


\bibitem[\protect\citeauthoryear{Wang, Ren, Mei, Chen, Ma, and de~Rijke}{Wang
  et~al\mbox{.}}{2019}]%
        {wang2019collaborative}
\bibfield{author}{\bibinfo{person}{Meirui Wang}, \bibinfo{person}{Pengjie Ren},
  \bibinfo{person}{Lei Mei}, \bibinfo{person}{Zhumin Chen},
  \bibinfo{person}{Jun Ma}, {and} \bibinfo{person}{Maarten de Rijke}.}
  \bibinfo{year}{2019}\natexlab{}.
\newblock \showarticletitle{A collaborative session-based recommendation
  approach with parallel memory modules}. In \bibinfo{booktitle}{\emph{The 42nd
  International ACM Conference on Research and Development in Information
  Retrieval}}. \bibinfo{pages}{345--354}.
\newblock


\bibitem[\protect\citeauthoryear{Wang, Fan, Xia, Zhao, Niu, and Huang}{Wang
  et~al\mbox{.}}{2020b}]%
        {wang2020kerl}
\bibfield{author}{\bibinfo{person}{Pengfei Wang}, \bibinfo{person}{Yu Fan},
  \bibinfo{person}{Long Xia}, \bibinfo{person}{Wayne~Xin Zhao},
  \bibinfo{person}{Shaozhang Niu}, {and} \bibinfo{person}{Jimmy Huang}.}
  \bibinfo{year}{2020}\natexlab{b}.
\newblock \showarticletitle{KERL: A knowledge-guided reinforcement learning
  model for sequential recommendation}. In \bibinfo{booktitle}{\emph{The 43rd
  International ACM Conference on Research and Development in Information
  Retrieval}}. \bibinfo{pages}{209--218}.
\newblock


\bibitem[\protect\citeauthoryear{Wang, Wei, Cong, Li, Mao, and Qiu}{Wang
  et~al\mbox{.}}{2020d}]%
        {wang2020global}
\bibfield{author}{\bibinfo{person}{Ziyang Wang}, \bibinfo{person}{Wei Wei},
  \bibinfo{person}{Gao Cong}, \bibinfo{person}{Xiao-Li Li},
  \bibinfo{person}{Xian-Ling Mao}, {and} \bibinfo{person}{Minghui Qiu}.}
  \bibinfo{year}{2020}\natexlab{d}.
\newblock \showarticletitle{Global context enhanced graph neural networks for
  session-based recommendation}. In \bibinfo{booktitle}{\emph{The 43rd
  International ACM Conference on Research and Development in Information
  Retrieval}}. \bibinfo{pages}{169--178}.
\newblock


\bibitem[\protect\citeauthoryear{Wu, Cai, and Wang}{Wu et~al\mbox{.}}{2020a}]%
        {wu2020deja}
\bibfield{author}{\bibinfo{person}{Jibang Wu}, \bibinfo{person}{Renqin Cai},
  {and} \bibinfo{person}{Hongning Wang}.} \bibinfo{year}{2020}\natexlab{a}.
\newblock \showarticletitle{D{\'e}j{\`a} vu: A contextualized temporal
  attention mechanism for sequential recommendation}. In
  \bibinfo{booktitle}{\emph{The 29th Web Conference}}.
  \bibinfo{pages}{2199--2209}.
\newblock


\bibitem[\protect\citeauthoryear{Wu, Wang, Feng, He, Chen, Lian, and Xie}{Wu
  et~al\mbox{.}}{2020c}]%
        {wu2020self}
\bibfield{author}{\bibinfo{person}{Jiancan Wu}, \bibinfo{person}{Xiang Wang},
  \bibinfo{person}{Fuli Feng}, \bibinfo{person}{Xiangnan He},
  \bibinfo{person}{Liang Chen}, \bibinfo{person}{Jianxun Lian}, {and}
  \bibinfo{person}{Xing Xie}.} \bibinfo{year}{2020}\natexlab{c}.
\newblock \showarticletitle{Self-supervised graph learning for recommendation}.
  In \bibinfo{booktitle}{\emph{arXiv preprint arXiv:2010.10783}}.
\newblock


\bibitem[\protect\citeauthoryear{Wu, Li, Hsieh, and Sharpnack}{Wu
  et~al\mbox{.}}{2020b}]%
        {wu2020sse}
\bibfield{author}{\bibinfo{person}{Liwei Wu}, \bibinfo{person}{Shuqing Li},
  \bibinfo{person}{Cho-Jui Hsieh}, {and} \bibinfo{person}{James Sharpnack}.}
  \bibinfo{year}{2020}\natexlab{b}.
\newblock \showarticletitle{SSE-PT: Sequential recommendation via personalized
  transformer}. In \bibinfo{booktitle}{\emph{The 14th ACM Conference on
  Recommender Systems}}. \bibinfo{pages}{328--337}.
\newblock


\bibitem[\protect\citeauthoryear{Wu, Li, Hsieh, and Sharpnack}{Wu
  et~al\mbox{.}}{2019a}]%
        {wu2019SSE}
\bibfield{author}{\bibinfo{person}{Liwei Wu}, \bibinfo{person}{Shuqing Li},
  \bibinfo{person}{Cho-Jui Hsieh}, {and} \bibinfo{person}{James~L Sharpnack}.}
  \bibinfo{year}{2019}\natexlab{a}.
\newblock \showarticletitle{Stochastic shared embeddings: data-driven
  regularization of embedding layers}. In \bibinfo{booktitle}{\emph{The 33rd
  Conference on Neural Information Processing Systems}}.
  \bibinfo{pages}{24--34}.
\newblock


\bibitem[\protect\citeauthoryear{Wu, Tang, Zhu, Wang, Xie, and Tan}{Wu
  et~al\mbox{.}}{2019b}]%
        {SRGNN}
\bibfield{author}{\bibinfo{person}{Shu Wu}, \bibinfo{person}{Yuyuan Tang},
  \bibinfo{person}{Yanqiao Zhu}, \bibinfo{person}{Liang Wang},
  \bibinfo{person}{Xing Xie}, {and} \bibinfo{person}{Tieniu Tan}.}
  \bibinfo{year}{2019}\natexlab{b}.
\newblock \showarticletitle{Session-based recommendation with graph neural
  networks}. In \bibinfo{booktitle}{\emph{The 33rd Conference on Artificial
  Intelligence}}. \bibinfo{pages}{346--353}.
\newblock


\bibitem[\protect\citeauthoryear{Xia, Yin, Yu, Wang, Cui, and Zhang}{Xia
  et~al\mbox{.}}{2021}]%
        {xia2020self}
\bibfield{author}{\bibinfo{person}{Xin Xia}, \bibinfo{person}{Hongzhi Yin},
  \bibinfo{person}{Junliang Yu}, \bibinfo{person}{Qinyong Wang},
  \bibinfo{person}{Lizhen Cui}, {and} \bibinfo{person}{Xiangliang Zhang}.}
  \bibinfo{year}{2021}\natexlab{}.
\newblock \showarticletitle{Self-supervised hypergraph convolutional networks
  for session-based recommendation}. In \bibinfo{booktitle}{\emph{The 35th
  Conference on Artificial Intelligence}}.
\newblock


\bibitem[\protect\citeauthoryear{Xie, Sun, Liu, Gao, Ding, and Cui}{Xie
  et~al\mbox{.}}{2020}]%
        {xie2020contrastive}
\bibfield{author}{\bibinfo{person}{Xu Xie}, \bibinfo{person}{Fei Sun},
  \bibinfo{person}{Zhaoyang Liu}, \bibinfo{person}{Jinyang Gao},
  \bibinfo{person}{Bolin Ding}, {and} \bibinfo{person}{Bin Cui}.}
  \bibinfo{year}{2020}\natexlab{}.
\newblock \showarticletitle{Contrastive Learning for Sequential
  Recommendation}.
\newblock \bibinfo{journal}{\emph{arXiv preprint arXiv:2010.14395}}
  (\bibinfo{year}{2020}).
\newblock


\bibitem[\protect\citeauthoryear{Xin, Karatzoglou, Arapakis, and Jose}{Xin
  et~al\mbox{.}}{2020}]%
        {xin2020self}
\bibfield{author}{\bibinfo{person}{Xin Xin}, \bibinfo{person}{Alexandros
  Karatzoglou}, \bibinfo{person}{Ioannis Arapakis}, {and}
  \bibinfo{person}{Joemon~M Jose}.} \bibinfo{year}{2020}\natexlab{}.
\newblock \showarticletitle{Self-supervised reinforcement learning for
  recommender systems}. In \bibinfo{booktitle}{\emph{The 43rd International ACM
  Conference on Research and Development in Information Retrieval}}.
  \bibinfo{pages}{931--940}.
\newblock


\bibitem[\protect\citeauthoryear{Yao, Yi, Cheng, Yu, Menon, Hong, Chi, Tjoa,
  Kang, and Ettinger}{Yao et~al\mbox{.}}{2020}]%
        {MIP2020}
\bibfield{author}{\bibinfo{person}{Tiansheng Yao}, \bibinfo{person}{Xinyang
  Yi}, \bibinfo{person}{Derek~Zhiyuan Cheng}, \bibinfo{person}{Felix~X. Yu},
  \bibinfo{person}{Aditya~Krishna Menon}, \bibinfo{person}{Lichan Hong},
  \bibinfo{person}{Ed~H. Chi}, \bibinfo{person}{Steve Tjoa},
  \bibinfo{person}{Jieqi Kang}, {and} \bibinfo{person}{Evan Ettinger}.}
  \bibinfo{year}{2020}\natexlab{}.
\newblock \showarticletitle{Self-supervised learning for deep models in
  recommendations}. In \bibinfo{booktitle}{\emph{arXiv preprint
  arXiv:2007.12865}}.
\newblock


\bibitem[\protect\citeauthoryear{Ye, Wang, Chen, Wang, Qin, and Yin}{Ye
  et~al\mbox{.}}{2020}]%
        {ye2020time}
\bibfield{author}{\bibinfo{person}{Wenwen Ye}, \bibinfo{person}{Shuaiqiang
  Wang}, \bibinfo{person}{Xu Chen}, \bibinfo{person}{Xuepeng Wang},
  \bibinfo{person}{Zheng Qin}, {and} \bibinfo{person}{Dawei Yin}.}
  \bibinfo{year}{2020}\natexlab{}.
\newblock \showarticletitle{Time matters: Sequential recommendation with
  complex temporal information}. In \bibinfo{booktitle}{\emph{The 43rd
  International ACM Conference on Research and Development in Information
  Retrieval}}. \bibinfo{pages}{1459--1468}.
\newblock


\bibitem[\protect\citeauthoryear{Yu, Zhu, Liu, Wu, Wang, and Tan}{Yu
  et~al\mbox{.}}{2020}]%
        {yu2020tagnn}
\bibfield{author}{\bibinfo{person}{Feng Yu}, \bibinfo{person}{Yanqiao Zhu},
  \bibinfo{person}{Qiang Liu}, \bibinfo{person}{Shu Wu}, \bibinfo{person}{Liang
  Wang}, {and} \bibinfo{person}{Tieniu Tan}.} \bibinfo{year}{2020}\natexlab{}.
\newblock \showarticletitle{TAGNN: Target attentive graph neural networks for
  session-based recommendation}. In \bibinfo{booktitle}{\emph{The 43rd
  International ACM Conference on Research and Development in Information
  Retrieval}}. \bibinfo{pages}{1921--1924}.
\newblock


\bibitem[\protect\citeauthoryear{Yu, Yin, Li, Wang, Hung, and Zhang}{Yu
  et~al\mbox{.}}{2021}]%
        {yu2021self}
\bibfield{author}{\bibinfo{person}{Junliang Yu}, \bibinfo{person}{Hongzhi Yin},
  \bibinfo{person}{Jundong Li}, \bibinfo{person}{Qinyong Wang},
  \bibinfo{person}{Nguyen Quoc~Viet Hung}, {and} \bibinfo{person}{Xiangliang
  Zhang}.} \bibinfo{year}{2021}\natexlab{}.
\newblock \showarticletitle{Self-supervised multi-channel hypergraph
  convolutional network for social recommendation}. In
  \bibinfo{booktitle}{\emph{arXiv preprint arXiv:2101.06448}}.
\newblock


\bibitem[\protect\citeauthoryear{Yuan, He, Alexandros, and Zhang}{Yuan
  et~al\mbox{.}}{2020a}]%
        {ssl_mask_peterrec2020}
\bibfield{author}{\bibinfo{person}{Fajie Yuan}, \bibinfo{person}{Xiangnan He},
  \bibinfo{person}{Karatzoglou Alexandros}, {and} \bibinfo{person}{Liguang
  Zhang}.} \bibinfo{year}{2020}\natexlab{a}.
\newblock \showarticletitle{Parameter-efficient transfer from sequential
  behaviors for user modeling and recommendation}. In
  \bibinfo{booktitle}{\emph{The 43rd International ACM Conference on Research
  and Development in Information Retrieval}}. \bibinfo{pages}{1469--1478}.
\newblock


\bibitem[\protect\citeauthoryear{Yuan, He, Jiang, Guo, Xiong, Xu, and
  Xiong}{Yuan et~al\mbox{.}}{2020b}]%
        {yuan2020future}
\bibfield{author}{\bibinfo{person}{Fajie Yuan}, \bibinfo{person}{Xiangnan He},
  \bibinfo{person}{Haochuan Jiang}, \bibinfo{person}{Guibing Guo},
  \bibinfo{person}{Jian Xiong}, \bibinfo{person}{Zhezhao Xu}, {and}
  \bibinfo{person}{Yilin Xiong}.} \bibinfo{year}{2020}\natexlab{b}.
\newblock \showarticletitle{Future data helps training: modeling future
  contexts for session-based recommendation}. In \bibinfo{booktitle}{\emph{The
  29th Web Conference}}. \bibinfo{pages}{303--313}.
\newblock


\bibitem[\protect\citeauthoryear{Zhai, Oliver, Kolesnikov, and Beyer}{Zhai
  et~al\mbox{.}}{2019}]%
        {cv_zhai2019}
\bibfield{author}{\bibinfo{person}{Xiaohua Zhai}, \bibinfo{person}{Avital
  Oliver}, \bibinfo{person}{Alexander Kolesnikov}, {and} \bibinfo{person}{Lucas
  Beyer}.} \bibinfo{year}{2019}\natexlab{}.
\newblock \showarticletitle{S4L: Self-supervised semi-supervised learning}. In
  \bibinfo{booktitle}{\emph{International Conference on Computer Vision}}.
  \bibinfo{pages}{1476--1485}.
\newblock


\bibitem[\protect\citeauthoryear{Zhang, Zhao, Liu, Sheng, Xu, Wang, Liu, and
  Zhou}{Zhang et~al\mbox{.}}{2019}]%
        {zhang2019feature}
\bibfield{author}{\bibinfo{person}{Tingting Zhang}, \bibinfo{person}{Pengpeng
  Zhao}, \bibinfo{person}{Yanchi Liu}, \bibinfo{person}{Victor Sheng},
  \bibinfo{person}{Jiajie Xu}, \bibinfo{person}{Deqing Wang},
  \bibinfo{person}{Guanfeng Liu}, {and} \bibinfo{person}{Xiaofang Zhou}.}
  \bibinfo{year}{2019}\natexlab{}.
\newblock \showarticletitle{Feature-level deeper self-attention network for
  sequential recommendation}. In \bibinfo{booktitle}{\emph{The 28th
  International Joint Conference on Artificial Intelligence}}.
  \bibinfo{pages}{4320--4326}.
\newblock


\bibitem[\protect\citeauthoryear{Zheng, Guo, Chen, Yu, and Jiang}{Zheng
  et~al\mbox{.}}{2020}]%
        {sentiSR2020}
\bibfield{author}{\bibinfo{person}{Lin Zheng}, \bibinfo{person}{Naicheng Guo},
  \bibinfo{person}{Weihao Chen}, \bibinfo{person}{Jin Yu}, {and}
  \bibinfo{person}{Dazhi Jiang}.} \bibinfo{year}{2020}\natexlab{}.
\newblock \showarticletitle{Sentiment-guided sequential recommendation}. In
  \bibinfo{booktitle}{\emph{The 43rd International ACM Conference on Research
  and Development in Information Retrieval}}. \bibinfo{pages}{1957--1960}.
\newblock


\bibitem[\protect\citeauthoryear{Zhou, Ma, Zhang, Zhou, and Yang}{Zhou
  et~al\mbox{.}}{2020a}]%
        {ssl_cl_queue2020}
\bibfield{author}{\bibinfo{person}{Chang Zhou}, \bibinfo{person}{Jianxin Ma},
  \bibinfo{person}{Jianwei Zhang}, \bibinfo{person}{Jingren Zhou}, {and}
  \bibinfo{person}{Hongxia Yang}.} \bibinfo{year}{2020}\natexlab{a}.
\newblock \showarticletitle{Contrastive learning for debiased candidate
  generation in large-scale recommender systems}. In
  \bibinfo{booktitle}{\emph{arXiv preprint arXiv:2005.12964}}.
\newblock


\bibitem[\protect\citeauthoryear{Zhou, Wang, Zhao, Zhu, Wang, Zhang, Wang, and
  Wen}{Zhou et~al\mbox{.}}{2020b}]%
        {ssl_cl_S3rec2020}
\bibfield{author}{\bibinfo{person}{Kun Zhou}, \bibinfo{person}{Hui Wang},
  \bibinfo{person}{WayneXin Zhao}, \bibinfo{person}{Yutao Zhu},
  \bibinfo{person}{Sirui Wang}, \bibinfo{person}{Fuzheng Zhang},
  \bibinfo{person}{Zhongyuan Wang}, {and} \bibinfo{person}{Ji{-}Rong Wen}.}
  \bibinfo{year}{2020}\natexlab{b}.
\newblock \showarticletitle{S3-Rec: Self-supervised learning for sequential
  recommendation with mutual information maximization}. In
  \bibinfo{booktitle}{\emph{The 29th Conference on Information and Knowledge
  Management}}. \bibinfo{pages}{1893--1902}.
\newblock


\bibitem[\protect\citeauthoryear{Zhuang, Zhou, Zhang, Ao, Xie, and He}{Zhuang
  et~al\mbox{.}}{2017}]%
        {cross-domain_novelty-seeking_sr}
\bibfield{author}{\bibinfo{person}{Fuzhen Zhuang}, \bibinfo{person}{Yingmin
  Zhou}, \bibinfo{person}{Fuzheng Zhang}, \bibinfo{person}{Xiang Ao},
  \bibinfo{person}{Xing Xie}, {and} \bibinfo{person}{Qing He}.}
  \bibinfo{year}{2017}\natexlab{}.
\newblock \showarticletitle{Cross-domain novelty seeking trait mining for
  sequential recommendation}. In \bibinfo{booktitle}{\emph{The 26th Web
  Conference}}. \bibinfo{pages}{881--882}.
\newblock


\end{thebibliography}

\end{document}